# A modified bond-based peridynamics model without limitations on elastic properties

Alireza Masoumi[1], Mohammad Ravandi[2], Manouchehr Salehi[1,*]

[1] *Department of Mechanical Engineering, Amirkabir University of Technology, Tehran 1591634311, Iran*

[2] *Department of Mechanical Engineering, K. N. Toosi University of Technology, Tehran 19919-43344, Iran*

*Corresponding author:

*Email address: msalehi@aut.ac.ir*

# A modified bond-based peridynamics model without limitations on elastic properties

**Abstract:**

This study proposes a novel Modified Bond-Based PeriDynamic (MBB-PD) model based on the bonds' classification. This classification of bonds is performed on the basis of the equivalent hypothetical local strains and falls into three categories of horizontal normal, vertical normal, and shear bonds. While the classical Bond-Based PD (BB-PD) considers only the stretch of bonds, all components of the bonds' strains are taken into account in the proposed model. A local imaginary element is considered around each bond to estimate the true strains of each bond. The constitutive relations are derived from equating the strain energies of the bonds' deformations to the Classical Continuum Mechanics (CCM) strain energies for a generalized combined loading condition. A novel critical stretch criterion and critical angle criterion are proposed to predict the failure of normal and shear strain bonds, respectively. It is also shown that, unlike the classical BB-PD, the proposed model does not impose any limitations on the value of Poisson's ratio. The model is verified by investigating some intact plane stress and plane strain problems under mechanical and thermal loadings. Moreover, the deformation and damage contours and the corresponding stress-strain responses are presented for different problems with pre-existing defects and validated with the eXtended Finite Element method's (XFEM) analysis.

**Nomenclature**

**Latin Letters**

| | |
|---|---|
| $A$ | The cross-sectional area of the structure |
| $a$, $b$, $c$, and $d$ | The dummy points on the edge of the hypothetical elements |
| $B$ | Continuum body |
| $b$ | Body force density |
| $[c_{ij}]$ | Generalized stiffness matrix of elasticity |
| $c$ | PD material constant |
| $c_1{}^t$, $c_2{}^t$ and $c_6{}^t$ | Summation of elements of the first, second, and third row of the material stiffness matrix |
| $E$ | Young's modulus of the material |
| $f$ | Force density |
| $|\vec{f}_{xy}^{\alpha}|$ | Magnitude of the shear bond force density vector in the $-\alpha$ type of bonds, |
| $G_M^C$ | Mixed-mode critical SERR |
| $G_\text{I}$ and $G_\text{II}$ | SEER of Mode I and Mode II deformations |
| $G_c$ | Critical SEER |
| $\mathcal{H}_x$ | Family region of a material point x |
| $h$ | Thickness of the structure |
| $K_1{}^t$, $K_2{}^t$ and $K_6{}^t$ | Summation of elements of the first, second, and third row of the micro-modulus matrix |
| $[K_{ij}]$ | Micro-modulus matrix |
| $k$ | Material's bulk modulus |
| $\vec{M}_{jk}$ | Unit vector of the j-k bond |
| $m$ | Ratio of the horizon size to the spacing size of the two adjacent points |
| $N_{(j)}^S$, $N_{(j)}^V$, and $N_{(j)}^H$ | Number of the equivalent shear bonds, the number of the equivalent vertical bonds and the number of the equivalent horizontal bonds of the material point of $j$ |
| $NF_{(j)}$ | Number of family members for material point j |
| $N_{cr}^V$, and $N_{cr}^S$ | Number of normal interactions, and the number of shear interactions which passed through a crack surface with the length of $\Delta x$ |
| $n$ | $n^{th}$ iteration |
| $REM_{(j)}^H$ | Number of existing horizontal bonds on the left or right side of the material point $j$ |
| $REM_{(j)}^V$ | Number of existing vertical bonds on the top or bottom side of the material point $j$ |
| $s_c$ | Critical stretch |
| $s_c^M$ and $\phi_c^M$ | Failure criterion for the normal bonds and shear bonds |
| $s_{jk}$, $s_{jk}{}^T$, and $\bar{s}_{jk}$ | Mechanical, thermal stretch, and the overall stretch |
| $S_\alpha^H$, $\alpha \in \{xx, yy, xy\}$ | Normal axial, transverse, and shear true strain components of an EHNSB |
| $S_\alpha^V$, $\alpha \in \{xx, yy, xy\}$ | Normal axial, transverse, and shear true strain components of an EVNSB |
| $S_\alpha^S$, $\alpha \in \{xx, yy, xy\}$ | Normal horizontal true strain, normal vertical true strain, and shear true strain of an ESSB |
| $t$ | Time |
| $U, \ddot{U}$ | Displacement and acceleration vectors |
| $U_{(\beta)}^\alpha$ | $\alpha$ component of the displacement of the material point $\beta$ |
| $U_{jk}^x$ and $U_{jk}^y$ | Relative displacement components of two material points $j$ and $k$, in the $x$ and $y$ directions |
| $V_{(k)}$ | Volume of material point k |
| $X$ | Initial position of a point |
| $Y^{\alpha}_{(\beta)}$ | $\alpha$ component of the deformed position of the material point $\beta$ |
| $Y$ | Deformed position of a point |

**Greek Letters**

| | |
|---|---|
| $\alpha$ | Thermal expansion coefficient |
| $\Delta T$ | Temperature changing |
| $\Delta x$ and $\Delta y$ | Spacing between two adjacent material points in $x$ and $y$ directions |
| $\delta$ | Horizon size |
| $\eta_{jk}$ | Relative displacement between two points of j and k |
| $\theta_{jk}$ | Angle of the shear bond |
| $\mu$ | Shear modulus |
| $\mu_{jk}$ | Function to represent state of interaction |
| $\nu$ | Value of Poisson's ratio |
| $\xi_0$ | Initial bond length of the smallest shear bond in the set of shear bonds with identical angles |
| $\xi_{jk}$ | Initial relative position between two points of j and k |
| $\rho$ | Mass density |
| $\sigma_{ij}$ and $\varepsilon_{ij}$ | In-plane stress and strain components |
| $\varphi_j$ | local damage index |
| $[\psi_{ij}]$ | Matrix of correction factors |

**Acronyms**

| | |
|---|---|
| ADR | Adaptive Dynamic Relaxation |
| ALM | Atomistic Lattice Models |
| BB-PD | Bond-Based PeriDynamic |
| B-K | Benzeggagh-Kenane |
| CCM | Classical Continuum Mechanics |
| CZM | Cohesive Zone Model |
| DCM | Discrete crack model |
| EHNSB | Equivalent Horizontal Normal Strain Bond |
| ESSB | Equivalent Shear Strain Bond |
| EVNSB | Equivalent Vertical Normal Strain Bond |
| FE | Finite Element |
| FNM | Floating Node method |
| LEFM | Linear Elastic Fracture Mechanics |
| MBB-PD | Modified Bond-Based PeriDynamic |
| MD | Molecular Dynamics |
| NOSB | Non-Ordinary State-Based |
| OSB | Ordinary State-Based |
| PD | Peridynamics |
| PF | Phase-Field |
| SCM | Smeared Crack Model |
| SERR | Strain Energy Release Rate |
| VCCT | Virtual Crack Closure Technique |
| XBEM | extended boundary element method |
| XFEM | eXtended Finite Element method's |

# 1. Introduction

The Finite Element (FE) method is a powerful tool for analyzing solid mechanics. Despite the high capabilities of this method, predicting damage initiation and propagation is usually a challenge. The governing equations in this method are undefined at discontinuities (e.g., cracks) because of the derivative essence of FE's fundamental equations. Therefore, it is essential to provide alternative methods to capture the failure behavior and damage propagation in structures. To date, various alternative methods have been undertaken to address the limitation of the CCM on damage prediction. Among them, PeriDynamics (PD) models are promising numerical tools in the analysis of imperfect structures, particularly those with notches, holes, and cracks.

A literature review has been compiled to categorize the main approaches in the development of failure analysis methods. These methods can be classified into three groups. The first group takes in "macro-to-micro" models, such as Virtual Crack Closure Technique (VCCT) [1,2], smeared crack model (SCM) [3,4], discrete crack model (DCM) [5,6], Cohesive Zone Model (CZM) [7,8], the eXtended Finite Element Method (XFEM) [9,10], Floating Node method (FNM) [11], and the extended boundary element method (XBFEM) [12,13]. The methods of this group are based on the Linear Elastic Fracture Mechanics (LEFM) and the Strain Energy Release Rate (SERR). Despite all the advantages in damage prediction, they have some difficulties in implementation and simulation. The main drawbacks of these approaches are the lack of characteristic lengths, the necessity of external damage growth criteria, and the demand for body re-meshing tasks [14].

The second group includes "micro-to-macro" approaches such as Molecular Dynamics Models (MDM) and Atomistic Lattice Models (ALM) [15,16]. Due to the singularity at crack tips in LEFM-based models, the MD in computational fracture mechanics applications has been specifically developed to study fracture behavior at singular points. MD is a nonlocal model in which all the model particles affect each other. In an MD simulation, the dynamics of atoms are modeled through balance equations of motion, which are discretized in time and integrated using a finite-difference algorithm to obtain the positions and velocities of the particles [17]. Then, stresses and strains at the atomic level and defects' dynamics can be easily determined in every location, including crack tips. The most significant problem with these models is their enormous computational cost, which makes modeling many realistic problems impossible on processors available today.

The third group comprises theories that link the above attitudes using particular scaling functions, such as the Phase-Field (PF) method [18–20] and the PeriDynamics (PD) theory [21,22]. Despite the extensive development of the PF, it is still plagued by some drawbacks, such as high computational cost and inaccuracy in predicting crack tip locations and crack thickness. Silling [21] introduced the theory of PD as an integral representation of the CCM relationships. Unlike the differential equations, describing governing equations by the integral formulations remains valid over discontinuities. The integral equations sum up the actual forces of the material points interacting with each other. This statement of the PD model is known as the bond-based (BB) model since material points interact through the bonds that connect two particles. Bonds in this model are analogous to springs since both sides of a bond have equal forces with inverse directions, governed by the laws of elasticity. On the other hand, in implementing PD equations, one can model the failure by checking the bonds' integrity at each deformation increment.

Consequently, the BB-PD model has been widely used to study the behavior of structures with discontinuities in many fields. For instance, many advances have been made in the analysis and prediction of static and dynamic fracture damage [23–27], impact failures [28–30], fatigue failures [31–33], and damages in composite laminates [23,25,34–37]. Nevertheless, the BB modeling of material points' relationships leads to two restrictions on interaction force vectors' freedoms in magnitudes and their directions. These force vector constraints impose specific limitations on the values of elastic properties. In other words, it can only model 2D and 3D problems with Poisson's ratios of $\frac{1}{3}$ and $\frac{1}{4}$, respectively [23].

In order to overcome Poisson's ratio limitations, various measures have been taken based on three strategies. The first strategy focused on the reduction of the force vector constraints. The mathematical "state" concept relax shortcomings in the PD's force vector modeling. Based on this concept, the Ordinary State-Based (OSB) [38,39] and the Non-Ordinary State-Based (NOSB) [40,41] PD models were introduced, which led to removing magnitude and direction restrictions of the force vectors. As a result of these modifications, the implementation of state-based PD models is generally more complicated and requires higher computational resources than the BB model [42].

The second strategy to eliminate Poisson's ratio limitations considered the rotational effect of material points. Previous studies demonstrated that the BB-PD model is limited by symmetrical characteristics of elasticity tensors [43]. This symmetry is due to the negligence of the couple-stress effects in the CCM theory [44]. Some continuum models consider the couple-stress effects, such as the micromorphic continuum theory [45,46] and the micropolar continuum theory [47,48]. Thus, the PD expression of micropolar continuum theory was developed in order to overcome the limitations of BB-PD [49–52]. This method, however, only transfers Poisson's ratio limitations to a range between $-1$ and $\frac{1}{3}$ [53].

The third modification made to the BB-PD model is the separation of motion equations depending on the bonds' directions. For instance, the tangential and normal deformations are used to derive the material points' equations of motion [54–57]. However, when Poisson's ratio equals $\frac{1}{3}$, the normal stiffness coefficient value is unexpectedly different from the BB-PD micro-modulus. In this regard, Zhu et al. [58] presented a truss element model of the particle bonds and derived the normal and tangential stiffness coefficients from categorizing strain energy functions. Despite the validity for intact isotropic materials under tension, their model has not been validated for predicting crack propagation and analysis of complex loading conditions [42]. Hue et al. [59–61] proposed a BB model based on bond classification by categorizing them into normal and shear bonds. The force density relationships and stiffness matrices are derived by comparing the CCM forces with the PD forces.

In this study, a novel modified BB-PD model is proposed to capture the elastic and damage response in isotropic mediums without the above limitations of the classical BB-PD. The BB-PD model is selected to modify because of its relatively lower computational cost than the other PD models. The bonds are classified as horizontal normal, vertical normal, and shear bonds to tackle Poisson's ratio limitation in the BB-PD, based on the equivalent strain bond behavior in a local imaginary element. Unlike the previous studies, the true strains are considered to increase the model's accuracy. The rest of this paper is organized as follows. Section 2 describes the BB-PD equations. Then, the strain and constitutive equations of the proposed model are presented in Section 3. Several numerical problems are discussed to investigate the validity and show the damage prediction capability of the proposed model in section 4. Finally, the conclusions are drawn in Section 5.

## 2. Bond-Based Peridynamic's equation of motion

The PD theory can be defined as the molecular dynamics' continuum model with a particular finite length scale called "horizon", $\delta$, [62]. In this representation of the continuum mechanics, as shown in Figure 1, each material point with the initial position $X$ in a continuum body $B$ has some family members with the initial position $X'$. All material points in the circle with the center of $X$ and the radius of $\delta$ are family members of $X$. Once the body is deformed, the new position of material points $X$ and $X'$ are $Y$ and $Y'$, respectively. The relationships between the positions of material points in the undeformed and deformed body can be written as:

$$Y = X + U$$
$$Y' = X' + U' \qquad (1)$$

where $U$ and $U'$ denote the displacement of $X$ and $X'$, respectively.

As described earlier, the PD equation is an integral representation of the CCM equation of motion, which is applied to all material points in the body as:

$$\rho \ddot{U}(X,t) = \int_{\mathcal{H}_x} f(U, U', X, X', t) dV_{X'} + b(X,t) \qquad (2)$$

where $\rho$, $\ddot{U}(X,t)$, and $b(X,t)$ are the mass density, the acceleration, and the body force density of the material point $X$, respectively. The integrand $f(U, U', X, X', t)$ is the force density acting on the bond between $X$ and $X'$. In Eq.(2), the integral domain $\mathcal{H}_x$ is the family region of each material point and is usually considered a circle, as shown in Figure 1.

Although some simple problems are solved analytically [63–65], the PD equations do not have generalized analytical solutions. As a result, numerical approaches are used to compute the spatial and time integrations of the PD equations. To do so, Eq.(2) is rewritten in the discretized form as:

$$\rho_{(j)}\ddot{U}_{(j)} = \sum_{k=1}^{NF_{(j)}} f_{jk}(\eta_{jk}, \xi_{jk}, t)V_{(k)} + b_{(j)} \quad (3)$$

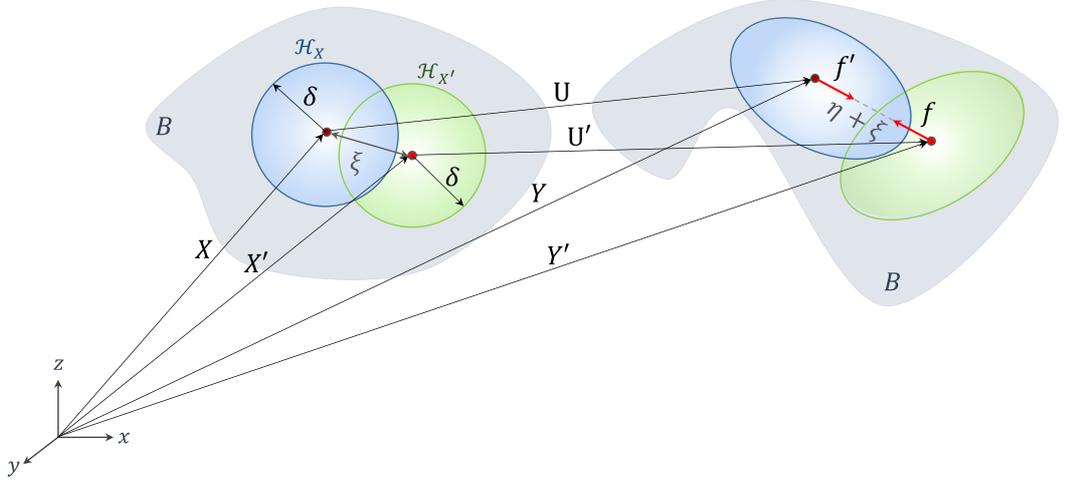

Figure 1. Material points' interactions in the BB-PD model.

where $\xi_{jk}$ and $\eta_{jk}$ are the bond length or the initial relative position ($\xi_{jk} = |X_{(j)} - X_{(k)}|$) and relative displacement ($\eta_{jk} = |U_{(j)} - U_{(k)}|$) of $j^{th}$ and $k^{th}$ material points, respectively, and $NF_{(j)}$ is the number of family members of $X_{(j)}$. In Eq.(3), $V_{(k)}$ and $b_{(j)}$ are the volume of the $k^{th}$ material point and the external body force for the $j^{th}$ material point, respectively. The force density relation $f_{jk}$ for a linear elastic isotropic solid body is given as:

$$f_{jk} = [c(\bar{s}_{jk} - s_{jk}^T)]\frac{Y_k - Y_j}{|Y_k - Y_j|} = cs_{jk}\vec{M}_{jk} \quad (4)$$

where $\vec{M}_{jk}$ is a unit vector in the direction of the linking bond from $X_{(j)}$ to $X_{(k)}$, $c$ is the material constant, $s_{jk}$, $s_{jk}^T$, and $\bar{s}_{jk}$, are the mechanical, thermal stretch, and the overall stretch, respectively, which are defined as:

$$s_{jk} = \frac{|Y_k - Y_j| - |X_k - X_j|}{|X_k - X_j|}$$

$$s_{jk}^T = \alpha \Delta T \quad (5)$$

where $\alpha$ and $\Delta T$ are the thermal expansion coefficient and the change in temperature, respectively. In the numerical implementation of Eq.(3), the true stretches can be used instead of the engineering stretches to enhance the accuracy of the basic BB-PD model. The true stretch can be defined as:

$$s_{jk}{}^n = \frac{\left|Y_k - Y_j\right|^n - \left|Y_k - Y_j\right|^{n-1}}{\left|Y_k - Y_j\right|^{n-1}} \tag{6}$$

where $n$ indicates the $n^{th}$ iteration. To determine the PD material constant $c$ in terms of the engineering material constants, the strain energy densities of a material point in the PD and CCM frameworks are set to be equal. This process is performed for two simple loading conditions, i.e., isotropic expansion and pure shear, which results in:

$$c = \frac{2E}{A\delta^2} \quad 1D$$

$$c = \frac{12k}{\pi h \delta^3} \quad 2D$$

$$c = \frac{18k}{\pi \delta^4} \quad 3D \tag{7}$$

In Eq.(7), $k$ is the material's bulk modulus, $E$ is Young's modulus, $h$ is the thickness, and $A$ is the cross-sectional area.

The failure parameter $\mu_{jk}$ is introduced to evaluate the damage effects on the mechanical response of the structure and incorporated into the equation of the motion as:

$$\rho_{(j)}\ddot{U}_{(j)} = \sum_{k=1}^{NF_{(j)}} \mu_{jk} f_{jk}\left(\eta_{jk}, \xi_{jk}, t\right) V_{(k)} + b_{(j)} \tag{8}$$

where $\mu_{jk}$ is defined as:

$$\mu_{jk} = \begin{cases} 1 & \text{if } \bar{s}_{jk} < s_c \text{ (bond exists)} \\ 0 & \text{if } \bar{s}_{jk} \geq s_c \text{ (bond breaks)} \end{cases} \tag{9}$$

in which $s_c$ is the critical stretch derived by equating the critical SERR of the material, $G_c$, to the corresponding strain energy density (SED) in the PD framework. At the moment of crack formation, the SED is equivalent to the work necessary to break all the bonds crossing the crack surface with a unit area. In an isotropic elastic body, $s_c$ is approximated as [62]:

$$s_c = \begin{cases} \sqrt{\dfrac{G_c}{\left(3\mu + \left(\dfrac{3}{4}\right)^4 \left(\kappa - \dfrac{5\mu}{3}\right)\right)\delta}} & \text{for 3D problems} \\ \sqrt{\dfrac{G_c}{\left(\dfrac{6}{\pi}\mu + \dfrac{16}{9\pi^2}(\kappa - 2\mu)\right)\delta}} & \text{for 2D problems} \end{cases} \quad (10)$$

where $\mu$ is the shear modulus. In addition, a local damage index is used to determine the status of damage propagation in the body, which is defined as the ratio of the broken bonds to the total number of bonds and expressed as follows:

$$\varphi_j = 1 - \frac{\sum_{k=1}^{NF(j)} \mu V_{(k)}}{\sum_{k=1}^{NF(j)} V_{(k)}} \quad (11)$$

where $\varphi_j$ is the local damage index corresponding to the material point $j$ and has a value between 0 and 1. Eq.(11) returns zero as long as all the bonds within the family of a material point are intact. If all the bonds between a material point and its family fail, then $\varphi_j = 1$.

## 3. Modified Bond-Based Peridynamic Model

As shown in Figure 2, family members of a material point in a 2D body experience three types of equivalent strain bonds: *Equivalent Horizontal Normal Strain Bond* (EHNSB), *Equivalent Vertical Normal Strain Bond* (EVNSB), and *Equivalent Shear Strain Bond* (EESSB). In the present study, the effects of these bonds are distinguished, and their force and stretch components in the governing equations are treated differently.

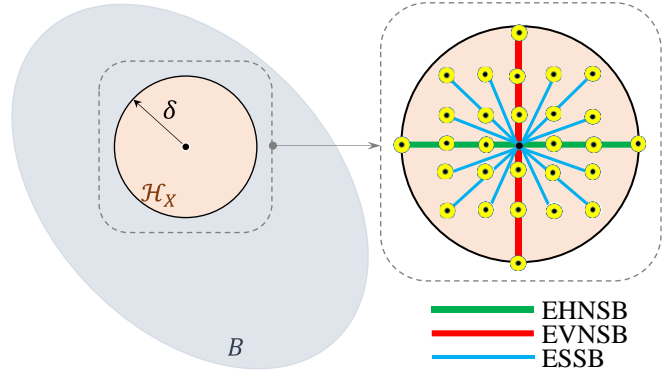

Figure 2. Classification of the equivalent strain bonds of a material point.

## 3.1. Local strain components

In this model, local imaginary elements are considered for each bond to obtain all the components of its deformation. Then, the strain components associated with these elements are derived. These strain components are then employed to compute the force densities of bonds.

*3.1.1. Equivalent Horizontal Normal Strain Bond*

As shown in Figure 3a, the interaction between $j$ and $k$ in the undeformed configuration is an EHNSB. The imaginary local rectangular element of $abdc$ is considered to capture the transformation of this bond after deformation. This element is formed by two adjacent points in the y-direction of $j$ (i.e., $a$ and $b$) and $k$ (i.e., $c$ and $d$).

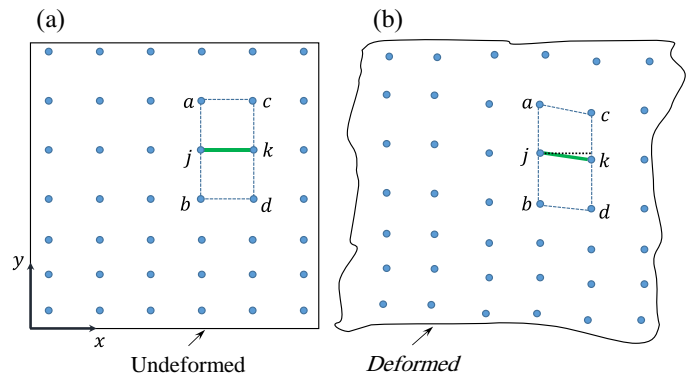

Figure 3. Local element and participating points due to an EHNSB; (a) undeformed body and (b) deformed body.

To enhance the accuracy in the numerical implementation of the fracture problems and avoid numerical instability, the true strains are employed instead of the engineering strains [66]. Therefore, under the infinitesimal strain consideration, the normal true strain associated with the EHNSB is approximated as:

$$S_{xx}^H = \frac{du}{dx} \approx \frac{U_{(j)}^x - U_{(k)}^x}{Y_{(j)}^x - Y_{(k)}^x} \tag{12}$$

As illustrated in Figure 3b, the corresponding element of the EHNSB in the deformed configuration has both the transverse and shear deformations. These true strains of the EHNSB can be estimated by:

$$S_{yy}^H = \frac{dv}{dy} \approx \frac{1}{2}\left(\frac{U_{(a)}^y - U_{(b)}^y}{Y_{(a)}^y - Y_{(b)}^y} + \frac{U_{(c)}^y - U_{(d)}^y}{Y_{(c)}^y - Y_{(d)}^y}\right)$$

$$S_{xy}^H = \frac{du}{dy} + \frac{dv}{dx}$$

$$\approx \frac{1}{2}\left\{\left(\frac{U_{(a)}^x - U_{(b)}^x}{Y_{(a)}^y - Y_{(b)}^y} + \frac{U_{(j)}^y - U_{(k)}^y}{Y_{(j)}^x - Y_{(k)}^x}\right)\right.$$

$$\left.+ \left(\frac{U_{(c)}^x - U_{(d)}^x}{Y_{(c)}^y - Y_{(d)}^y} + \frac{U_{(j)}^y - U_{(k)}^y}{Y_{(j)}^x - Y_{(k)}^x}\right)\right\} \tag{13}$$

where $S_\alpha^H, \alpha \in \{xx, yy, xy\}$ are the normal axial, transverse, and shear true strain components of an EHNSB. Furthermore, $U_{(\beta)}^\alpha$ and $Y_{(\beta)}^\alpha$ are the $\alpha$ component of the displacement and deformed position of the material point $\beta$, respectively.

*3.1.2. Equivalent Vertical Normal Strain Bond*

Similar to the EHNSB, the EVNSB's local imaginary element contains the vertical interaction between material points $j$ and $k$ in the undeformed configuration and their adjacent points in the x-direction (i.e., $a$, $b$, $c$, and $d$) are considered, as illustrated in Figure 4.

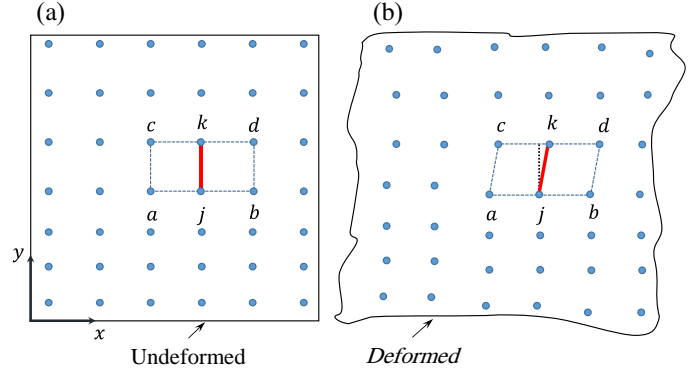

Figure 4. Imaginary local element and participating points due to an EVNSB; (a) undeformed body and (b) deformed body.

Thus, the true strain components of an EVNSB can be calculated as follows:

$$S_{xx}^V = \frac{dv}{dy} \approx \frac{U_{(j)}^y - U_{(k)}^y}{Y_{(j)}^y - Y_{(k)}^y}$$

$$S_{yy}^V = \frac{du}{dx} \approx \frac{1}{2}\left(\frac{U_{(a)}^x - U_{(b)}^x}{Y_{(a)}^x - Y_{(b)}^x} + \frac{U_{(c)}^x - U_{(d)}^x}{Y_{(c)}^x - Y_{(d)}^x}\right)$$

$$S_{xy}^V = \frac{du}{dy} + \frac{dv}{dx}$$

$$\approx \frac{1}{2}\left\{\left(\frac{U_{(j)}^x - U_{(k)}^x}{Y_{(j)}^y - Y_{(k)}^y} + \frac{U_{(a)}^y - U_{(b)}^y}{Y_{(a)}^x - Y_{(b)}^x}\right)\right.$$

$$\left. + \left(\frac{U_{(j)}^x - U_{(k)}^x}{Y_{(j)}^y - Y_{(k)}^y} + \frac{U_{(c)}^y - U_{(d)}^y}{Y_{(c)}^x - Y_{(d)}^x}\right)\right\} \quad (14)$$

where $S_\alpha^V, \alpha \in \{xx, yy, xy\}$ are the normal axial, transverse, and shear true strain components of an EVNSB.

*3.1.3. Equivalent Shear Strain Bond*

If the interaction between $j$ and $k$ in the undeformed configuration is neither an EHNSB nor an EVNSB, it is considered as an Equivalent Shear Strain Bond (ESSB). As indicated in Figure 5, the local imaginary rectangular element corresponding to the ESSB is formed based on the diagonal interaction between material point $j$ and $k$, and points $a$ and $b$ in the undeformed configuration. In this local element, the point $a$ has the same $x$ coordinate with the point $j$ and the same $y$ coordinate with the point $k$. Similarly, point $b$ has the same $x$ and $y$ coordinates with points $k$ and $j$, respectively.

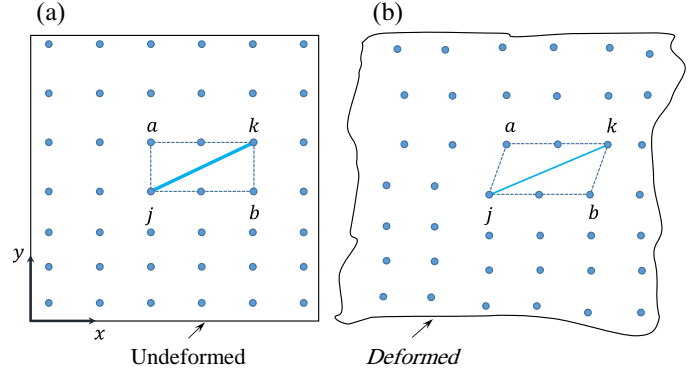

Figure 5. Local element and participating points due to an ESSB; (a) undeformed body and (b) deformed body.

Thus, the corresponding true strains of ESSB can be expressed as:

$$S_{xx}^s = \frac{du}{dx} \approx \frac{1}{2}\left(\frac{U_{(j)}^x - U_{(b)}^x}{Y_{(j)}^x - Y_{(b)}^x} + \frac{U_{(k)}^x - U_{(a)}^x}{Y_{(k)}^x - Y_{(a)}^x}\right)$$

$$S_{yy}^s = \frac{dv}{dy} \approx \frac{1}{2}\left(\frac{U_{(j)}^y - U_{(a)}^y}{Y_{(j)}^y - Y_{(a)}^y} + \frac{U_{(k)}^y - U_{(b)}^y}{Y_{(k)}^y - Y_{(b)}^y}\right)$$

$$S_{xy}^s = \frac{du}{dy} + \frac{dv}{dx}$$

$$\approx \frac{1}{2}\left\{\left(\frac{U_{(j)}^x - U_{(a)}^x}{Y_{(j)}^y - Y_{(a)}^y} + \frac{U_{(j)}^y - U_{(b)}^y}{Y_{(j)}^x - Y_{(b)}^x}\right) + \left(\frac{U_{(k)}^x - U_{(b)}^x}{Y_{(k)}^y - Y_{(b)}^y} + \frac{U_{(j)}^y - U_{(a)}^y}{Y_{(j)}^x - Y_{(a)}^x}\right)\right\} \quad (15)$$

where $S_\alpha^s, \alpha \in \{xx, yy, xy\}$ are the normal horizontal true strain, normal vertical true strain, and shear true strain of an ESSB.

### 3.2. The force density-strain relations

The bond force densities are determined by multiplying the bond stretch by the micro-modulus, as in the BB-PD model. However, in the proposed model, we have to distinguish the micro-modulus values corresponding to each type of strain described in the previous section. This leads to a matrix form for the micro-modulus. As a result, the force density-strain relationship is expressed as:

$$\begin{Bmatrix} f_{xx}^\alpha \\ f_{yy}^\alpha \\ f_{xy}^\alpha \end{Bmatrix} = \begin{bmatrix} K_{11} & K_{12} & K_{16} \\ K_{12} & K_{22} & K_{26} \\ K_{16} & K_{26} & K_{66} \end{bmatrix} \begin{Bmatrix} S_{xx}^\alpha \\ S_{yy}^\alpha \\ S_{xy}^\alpha \end{Bmatrix} \qquad (16)$$

According to Figure 6, the force density of the shear bond can be expressed in the vector form as:

$$\vec{f}_{xy}^\alpha = \frac{1}{2}|f_{xy}^\alpha|\left\{\left(\frac{\xi}{\xi_0}\tan\theta_{jk}\right)\vec{\imath} + \left(\frac{\xi}{\xi_0}\cot\theta_{jk}\right)\vec{\jmath}\right\} \qquad (17)$$

where $\vec{\imath}$ and $\vec{\jmath}$ are the unit vector in the $x$ and $y$ directions of the Cartesian coordinate system. Furthermore, $|\vec{f}_{xy}^\alpha|$ is the magnitude of the shear bond force density vector in the $-\alpha$ type of bonds, which can be the vertical, horizontal, and shear bonds. Here, $\theta_{jk}$ is the angle of the shear bond with respect to the $x$-axis. Furthermore, $\xi$ and $\xi_0$ are the current and initial bond length of the smallest shear bond in the set of shear bonds with identical angles, respectively.

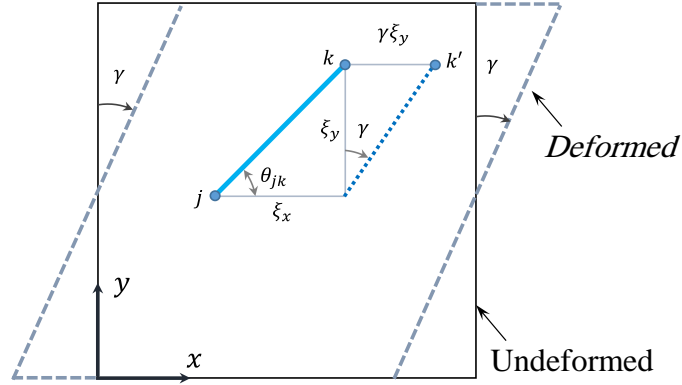

Figure 6. The relative position of two material points in a simple shear loading.

The strain energy calculated from the PD is set to be equal to the corresponding strain energy from the CCM under a combined loading condition, including biaxial and shear loads, to obtain the micro-modulus matrix. This process is described in detail in Appendix A. The micro-modulus matrix is given as:

$$\begin{bmatrix} K_{11} & K_{12} & K_{16} \\ K_{12} & K_{22} & K_{26} \\ K_{16} & K_{26} & K_{66} \end{bmatrix} = \begin{bmatrix} \psi_{11} & \psi_{12} & \psi_{16} \\ \psi_{21} & \psi_{22} & \psi_{26} \\ \psi_{61} & \psi_{62} & \psi_{66} \end{bmatrix} \begin{bmatrix} c_{11} & c_{12} & c_{16} \\ c_{12} & c_{22} & c_{26} \\ c_{16} & c_{26} & c_{66} \end{bmatrix} \qquad (18)$$

where $[c_{ij}]$ is the generalized stiffness matrix of elasticity, and $[\psi_{ij}]$ is the matrix of correction factors. By equating the PD and CCM strain energies, the elements of the correction factor matrix are calculated as:

$$\psi_{11} = \psi_{12} = \psi_{16}$$

$$= 2V_{(j)} \left( m(m+1)\Delta x + \sum_{k=1}^{N_{(j)}^S} \left( \left(1 + \frac{1}{2}\tan\theta_{jk}\right)^2 |\vec{\xi}_{jk}| \right) \right)^{-1}$$

$$\psi_{21} = \psi_{22} = \psi_{26}$$

$$= 2V_{(j)} \left( m(m+1)\Delta y + \sum_{k=1}^{N_{(j)}^S} \left( \left(1 + \frac{1}{2}\cot\theta_{jk}\right)^2 |\vec{\xi}_{jk}| \right) \right)^{-1}$$

$$\psi_{61} = \psi_{62} = \psi_{66} = 2V_{(j)} \left( \sum_{k=1}^{N_{(j)}^S} \left( (\csc 2\theta_{jk} + 1)^2 |\vec{\xi}_{jk}| \right) \right)^{-1} \qquad (19)$$

where $\Delta x$ and $\Delta y$ are the spacing between two adjacent material points in $x$ and $y$ directions, respectively, and $m = \frac{\delta}{\Delta x}$. These values have been derived for material points, which are far from free surfaces and have a complete horizon. In the regions close to the boundaries, some interactions are lost due to the attenuation of the family members. Some surface correction factors to address these inaccuracies are indispensable. Detailed explanations about deriving the surface correction factors can be found in Appendix B.

### 3.3. Failure criteria

Similar to the BB-PD model, to capture the damage growth, a failure parameter is introduced into Eq.(2) as follows:

$$\mu = \begin{cases} 1 & \text{if interaction exists} \\ 0 & \text{if interaction breaks} \end{cases} \qquad (20)$$

The failure parameter $\mu$ determines the status of an interaction. If the failure criterion is met, the bond interaction is broken, and the associated bond force becomes zero. Otherwise, the failure parameter is one. The failure criterion is presented for the mixed-mode I and II, which can be switched to each mode easily based on the SEER of each deformation, as shown in Figure 7.

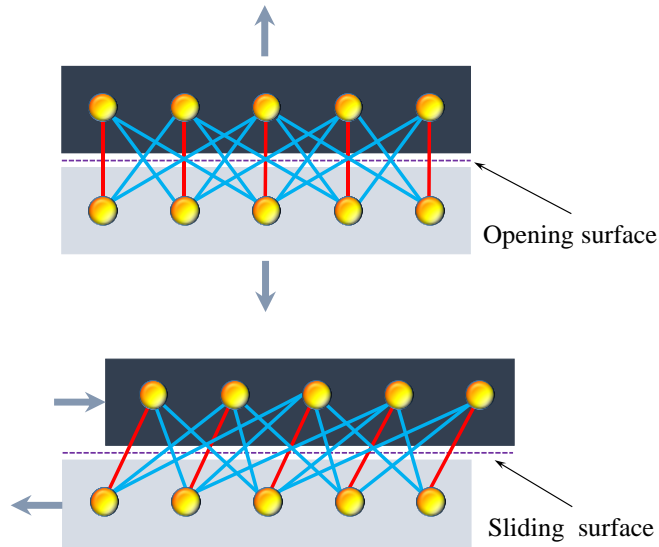

Figure 7. Material points' deformation in Modes I and II of failure.

The failure criterion for the normal bonds and shear bonds are $s_c^M$ and $\phi_c^M$, respectively. They are defined based on the mixed-mode critical SERR ($G_M^C$), which is approximated using the Benzeggagh-Kenane (B-K) formula [67]:

$$G_M^C = G_{IC} + (G_{IIC} - G_{IC})\left(\frac{G_{II}}{G_T}\right)^\eta, \text{ with } G_T = G_I + G_{II} \tag{21}$$

where $G_I$ and $G_{II}$ are the SEER of Mode I and Mode II deformations at each time step, and $\eta$ is the B-K constant. Also, the SEERs for a material point can be approximated from:

$$G_{I(j)} = \sum_{k=1}^{NF_{(j)}} \frac{\mu f_{yy}^\alpha S_{yy}^\alpha U_{jk}^y V_{(j)} V_{(k)}}{2(\Delta x)h}$$

$$G_{II(j)} = \sum_{k=1}^{NF_{(j)}} \frac{\mu f_{xx}^\alpha S_{xx}^\alpha U_{jk}^x V_{(j)} V_{(k)}}{2(\Delta x)h} \tag{22}$$

where $U_{jk}^x$ and $U_{jk}^y$ are relative displacement components of two material points $j$ and $k$, in the $x$ and $y$ directions, respectively. Furthermore, $\alpha$ indicates the type of bonds, which can be vertical, horizontal, or shear bonds. Consequently, the $\phi_c^M$ and $s_c^M$ values can be defined as:

$$s_{c(j)}^M = \left(\left(\sum_{k=1}^{N_{cr}^V} \mu\left\{\left(K_1^{\ t} + \frac{1}{2}\cot\theta_{jk} K_6^{\ t}\right)\vec{j}\right\}\vec{\xi}_{jk}\right.\right.$$

$$+ \sum_{k=1}^{N_{cr}^S} \mu\left\{\left((\tan\theta_{jk} - \nu_{xy})^2 K_1^{\ t} + \frac{1}{2}\tan\theta_{jk}(1+\sin\theta_{jk})^2 K_6^{\ t}\right)\vec{\iota}\right.$$

$$\left.\left. + \left(K_2^{\ t}\vec{j} + \frac{1}{2}\cot\theta_{jk}(1+\sin\theta_{jk})^2 K_6^{\ t}\right)\vec{j}\right\}\vec{\xi}_{jk}\right)^{-1} 2G_M^C \Delta x h V^{-2}\right)^{\frac{1}{2}} \tag{23}$$

in which the critical angle $\phi_{c(jk)}^M$ is:

$$\phi_{c(jk)}^M = s_{c(j)}^{II}(\pm 1 \pm \sin\theta_{jk}) \tag{24}$$

The signs of the sinusoidal term and 1 in the critical angle relation depend on the shear and normal loading signs, respectively. Detailed derivations of these failure criterion parameters are described in Appendix C. In Eq.(23), $K_1^t$, $K_2^t$ and $K_6^t$ are the summation of elements of the first, second, and third row of the micro-modulus matrix, respectively. Furthermore, $h$, $N_{cr}^V$, and $N_{cr}^S$ are the thickness of structure, the number of normal interactions, and the number of shear interactions which passed through a crack surface with the length of $\Delta x$ (Figure 8). For instance, in a case with $\delta = 3\Delta x$, the number of normal and shear interactions which pass through a crack surface with length $\Delta x$ is 12 and 32, respectively.

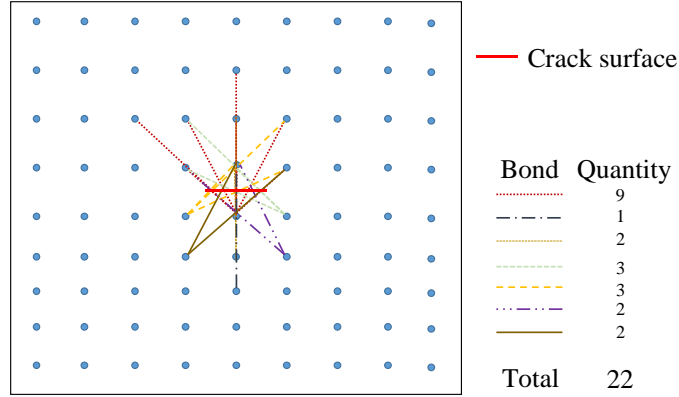

Figure 8. Material points interact with each other around a crack surface with a length of $\Delta x$.

The local damage index $\varphi_j$ is introduced to manifest the damage propagation status. The index $\varphi_j$ is defined as the ratio of the broken interactions to the total number of interactions corresponding to a material point within its horizon as:

$$\varphi_j = 1 - \frac{\sum_{k=1}^{NF(j)} \mu V_{(k)}}{\sum_{k=1}^{NF(j)} V_{(k)}} \tag{25}$$

The local damage index has a value between 0 and 1. $\varphi_j = 0$ indicates that all the bonds are intact, and $\phi_j = 1$ implies that all the interactions of material point $j$ are broken.

# 4. Numerical results

In this section, we first demonstrate the verification of the proposed model by comparing the responses of isotropic intact structures for plane stress and plane strain problems and different mechanical and thermal boundary conditions. Then, damage propagation in plane stress and plane strain problems with pre-existing defects and under general loading conditions is investigated. All problems are assumed to have the same elastic modulus of $E = 65.8\ GPa$. However, the Poisson's ratios are altered for each problem to indicate the accuracy and validity of the MBB-PD model in all Poisson's ratios.

## 4.1. Plane stress model

In the first problem, a plate structure under different loading conditions is investigated. In these problems, the plane stress material model is employed. The plane stress stiffness matrix is given as:

$$c = \frac{E}{1-v^2} \begin{bmatrix} 1 & v & 0 \\ v & 1 & 0 \\ 0 & 0 & \frac{1-v}{2} \end{bmatrix} \quad (26)$$

A square plate of length $100\ mm$ and thickness of $1\ mm$ is considered for analysis. The PD modeling of this problem includes a single layer of material points with a grid size of $\Delta x = \Delta y = 0.5\ mm$. In order to establish spatial integration, a circular domain of interaction is considered with a horizon size of $\delta = 3\Delta x$. For numerical time integration, the Adaptive Dynamic Relaxation (ADR) method is utilized with a time step value of one.

### 4.1.1. Plate under mechanical loading

In this section, the plate is subjected to uniaxial tensile loading in horizontal and vertical directions, and a shear displacement, as shown in Figure 9(a) to (c). In the uniaxial tension cases, the plate's top and bottom material points are subjected to a tensile loading of $p = 200\ MPa$. This problem is solved for Poisson's ratios of $v = -0.4,\ -0.2, 0, 0.2$ and $0.4$. These cases are repeated with horizontal loading applied on the left and right sides of the plate. The shear behavior of the plate is investigated by applying a horizontal displacement of $U^x = 1\ mm$ on the top boundary while constraining the displacement on the bottom boundary (as shown in Figure 9(c)).

Figure 9(d)-(f) show the deformation contours along the loading directions with respect to Figure 9(a) to (c). As expected, the plates deform in a uniform, symmetrical manner under these simple loading conditions. Furthermore, the transverse deformations of point A are compared with the analytical results for a range of Poisson's ratios in Figure 9(g)-(i). Compared with the exact solutions, it is evident that the MBB-PD results are highly accurate for all Poisson's ratios and loading conditions.

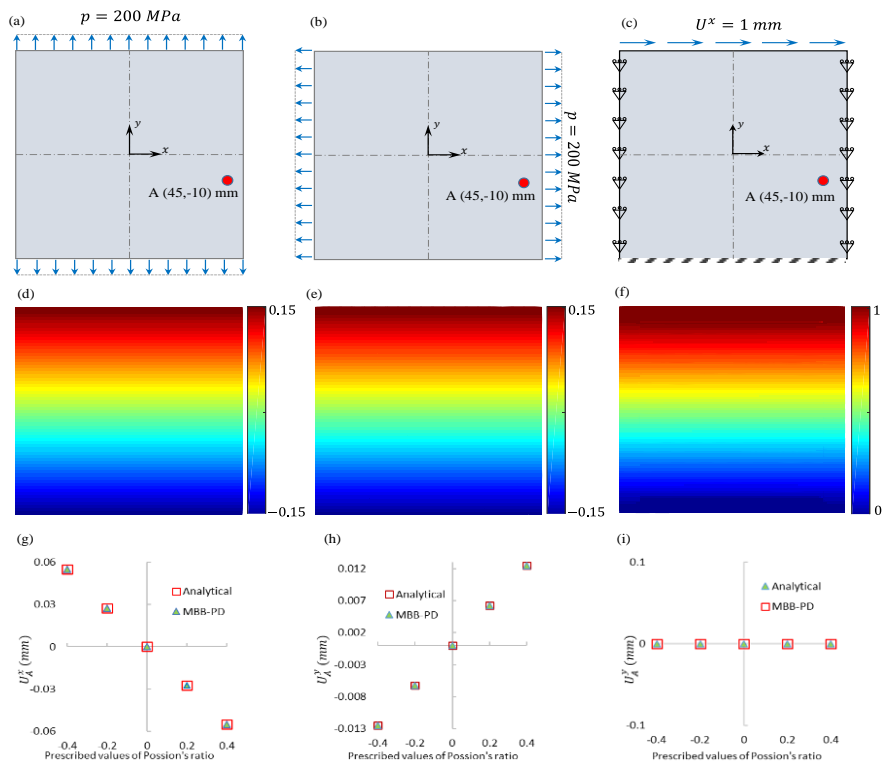

Figure 9- Schematic figures and the results of the plate under mechanical loadings. The schematic figures of the uniaxial tensile loading in the y-direction (a), the uniaxial tensile in the x-direction (b), and the shear loading (c). (d) to (f) illustrate deformation contours of the plate corresponding to the loading conditions of (a) to (c). (d) to (f) compare the MBB-PD and the analytical results for the transverse deformation of point A corresponding to the loading conditions of (a) to (c) for different Poisson's ratios.

*4.1.2. Plate under thermal loading*

The validation of the MBB-PD model is continued by applying a uniform temperature change of $\Delta T = 50 \,°C$ over the plate with a Poisson's ratio of $v = 0.4$. Two types of mechanical boundary conditions are investigated. In the first case, only the top and bottom boundaries are clamped, and the right and left boundaries are free, as shown in Figure 10(a). The horizontal and vertical deformation contours are compared with the FEM results in Figure 10(b) to (e). It can be observed that there is a good agreement between FE and MBB-PD results, and the MBB-PD model can capture the thermoelastic deformation with reasonable accuracy.

To validate the model's ability to capture more complex deformations, the previous problem is repeated with an additional clamped boundary condition on the left side of the plate, as shown in Figure 10(f). Figure 10(g) to (j) show that both the horizontal and vertical displacement contours of the MBB-PD model are well matched to those obtained from FEM. Also, the displacement at the free boundary is well captured.

**4.2. Plane strain model**

The MBB-PD model introduced in this paper is a generalized model that one can utilize to solve plane strain problems. Here, a long pressure vessel with internal pressure loading conditions is modeled to demonstrate the validity of the model for plane strain problems. The stiffness matrix of the plane strain material model is given as:

$$C = \frac{E}{(1+v)(1-2v)} \begin{bmatrix} 1-v & v & 0 \\ v & 1-v & 0 \\ 0 & 0 & \frac{1-2v}{2} \end{bmatrix} \qquad (27)$$

A long pressure vessel is selected with an internal radius of $r_i = 1.5 \, m$ and an external radius of $r_o = 2.2 \, m$. The polar coordinate system's location is in the vessel's center. The numerical model is constructed with a single layer of material points with a grid size of $\Delta r = \Delta \theta = 0.001 m$. The horizon size and numerical method are the same as the plane stress problems.

*4.2.1. Cylinder under uniform internal pressure*

In this problem, the long vessel with plane strain conditions is subjected only to an internal pressure loading $p = 400\ MPa$, as shown in Figure 11(a). This problem is solved for two different Poisson's ratios of $v = -0.1$ and $0.1$. The radial displacement counters for each Poisson's ratio obtained from the MBB-PD model and FEM are demonstrated in Figure 11(b) to (e). It can be seen that all contours have an identical range of displacements. Also, by varying the sign of Poisson's ratio, the displacements are alternated reciprocally. These results indicate the accuracy and validity of the MBB-PD model for both the plane stress and plane strain problems.

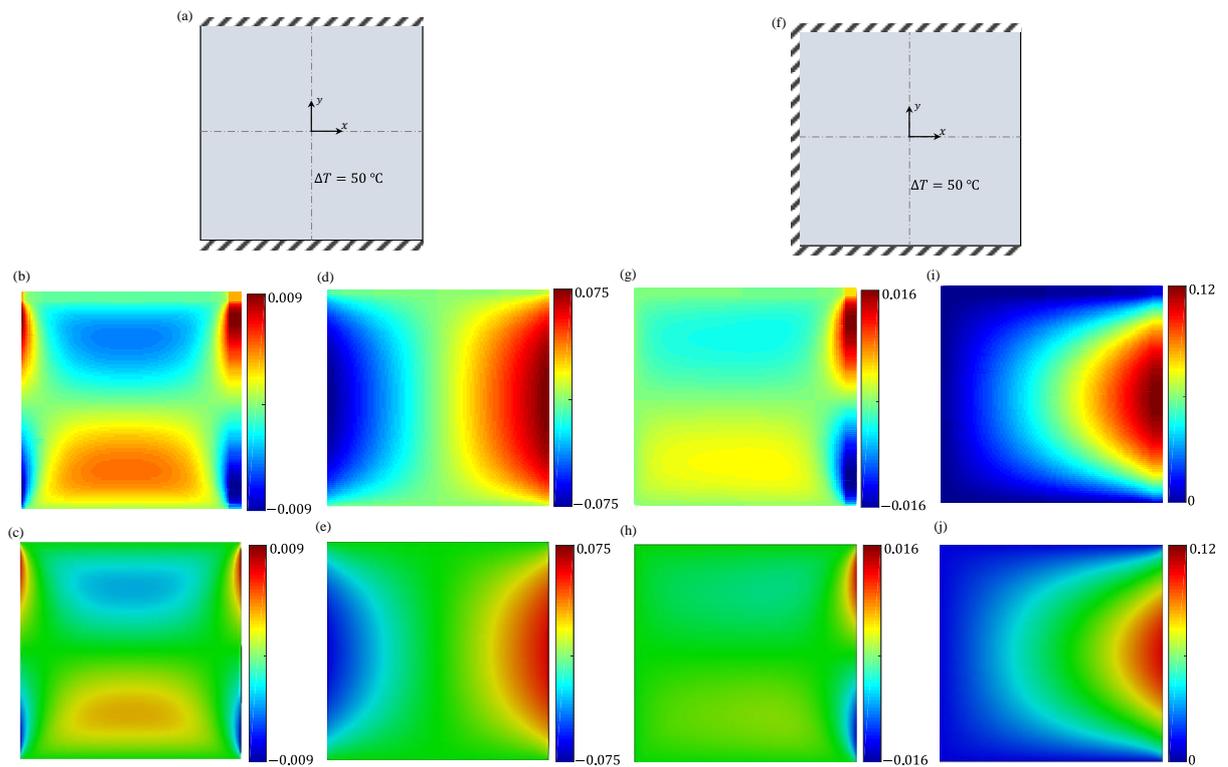

Figure 10- Schematic figures and deformation contours of the plate under uniform temperature change. Plate with two clamped and two free edges, (a) dimensions and boundary conditions, (b) and (c) MBB-PD and FEM contours of the vertical displacements, (d) and (e) MBB-PD and FEM contours of the horizontal displacements. Plate with three clamped and one free edge, (f) dimensions and boundary conditions, (g) and (h) MBB-PD and FEM contours of the vertical displacements, (i) and (j) MBB-PD and FEM contours of the horizontal displacements.

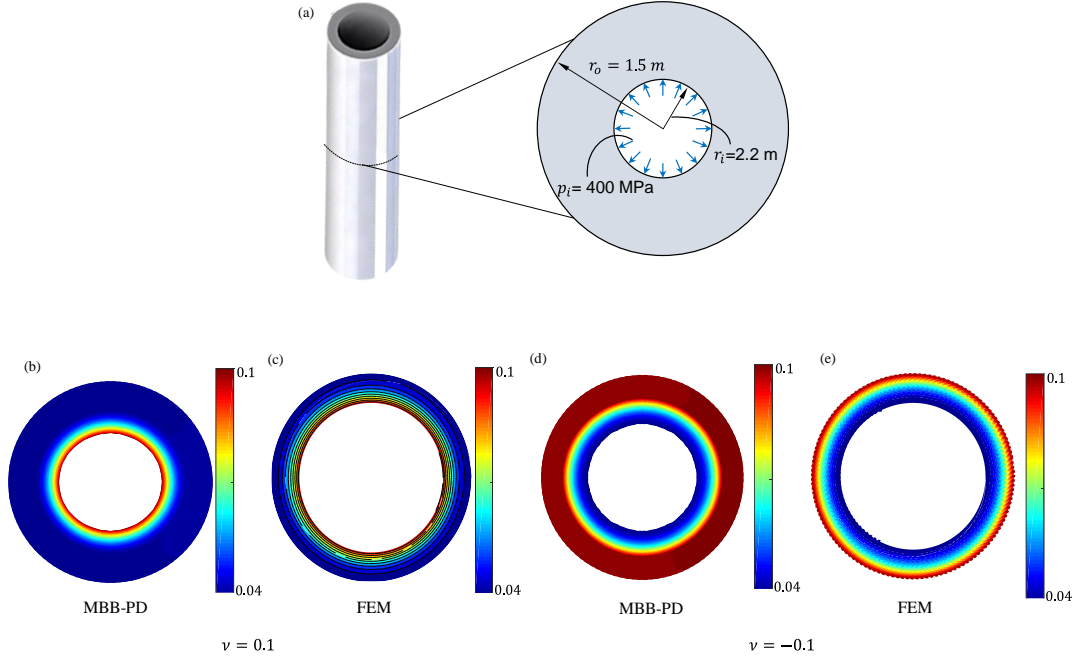

Figure 11- Cylinder under uniform internal pressure, (a) dimensions and conditions of the problem, radial displacement for $v = 0.1$ based on the (b) MBB-PD and (c) FEM, and (d) MBB-PD and (e) FEM contours for $v = -0.1$.

### 4.3. Failure prediction

Several problems with pre-existing defects are investigated to demonstrate the failure prediction capability of the MBB-PD method. Material properties are the same as in previous sections and with the Poisson's ratio of $v = 0.3$. The critical SERR of the material is $G_c = 6137 \text{ N/m}^2$. Furthermore, dimensions, grid size, horizon size, and numerical procedure are the same as in section 4.1.1. All failure problems have a tensile displacement of $1 \, mm$ on the top boundary along the $y$-direction and the vertical displacement component of the bottom edge is set to be zero.

In the first example, the damage behavior of a plate containing a central horizontal crack with a length of $0.01 \, m$ (see Figure 12) is investigated. The stress-strain curve obtained from the MBB-PD is given in Figure 13 and compared with the XFEM simulation result. As can be seen, the MBB-PD model shows reasonably good agreement with the results of the XFEM analysis. Although the peak load is slightly different, the predicted response in terms of the linear behavior of the stress-strain curve and the failure strain matches the XFEM results. The structure's load-bearing capacity is sharply dropped in the crack propagation phase. The difference between curves after the peak load point arises from the degradation models and the numerical methods.

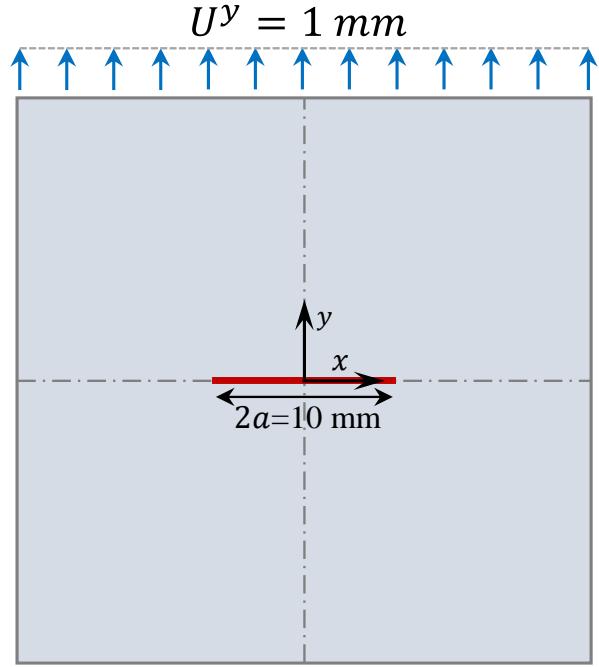

Figure 12- Dimensions and boundary conditions of the plate with a pre-existing central crack under tensile displacement.

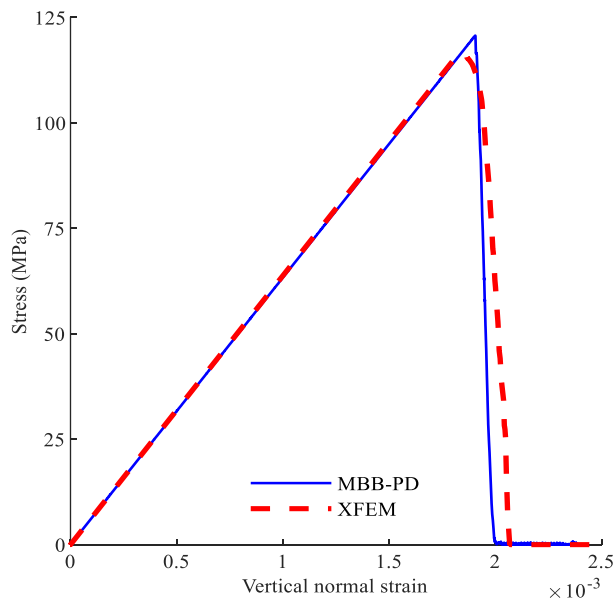

Figure 13- Comparison of the stress-strain responses of the MBB-PD and XFEM for the plate with a pre-existing central crack under tensile loading.

Figure 14 compares vertical displacement contours in MBB-PD and XFEM models in different time steps during crack propagation. A close resemblance can be observed between the vertical deformations obtained from XFEM and MBB-PD. Furthermore, the crack propagates symmetrically along a straight horizontal line in both models, as expected.

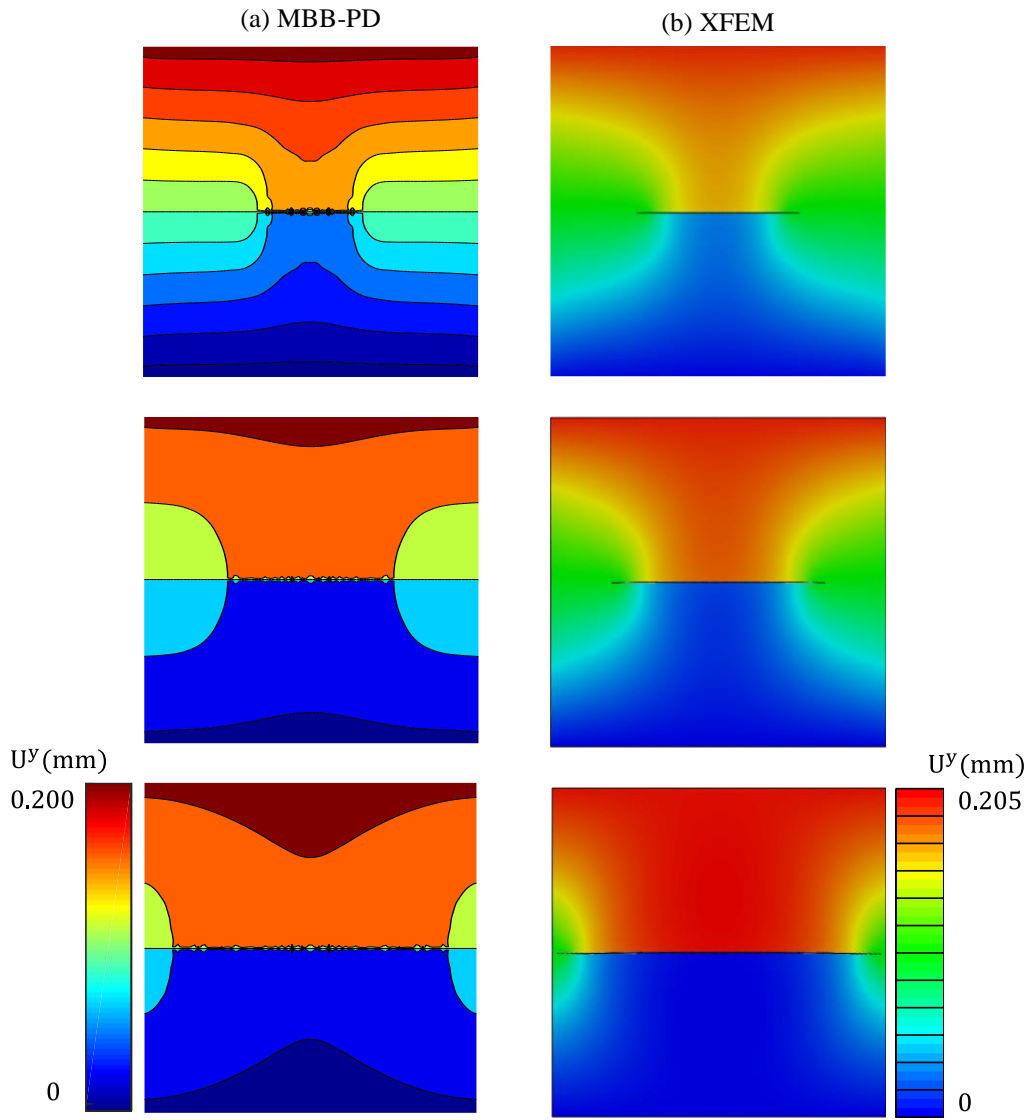

Figure 14- The vertical deformation contours of the plate with a pre-existing central crack subjected to the tensile loading in various time steps, which are derived based on the (a) the MBB-PD model and (b) XFEM.

In the second example, the MBB-PD model is employed to demonstrate its validity in predicting the failure of a plate with multiple cracks. The plate has three cracks with an identical length of $0.01\ m$. These cracks are located at the middle of the right, left, and top edge of the plate, as shown in Figure 15(a). In the last example, a plate having a square hole of $0.01 \times 0.01\ mm$ in the middle is analysed (see Figure 15(b)). Both cases have similar boundary conditions to that considered in the first example.

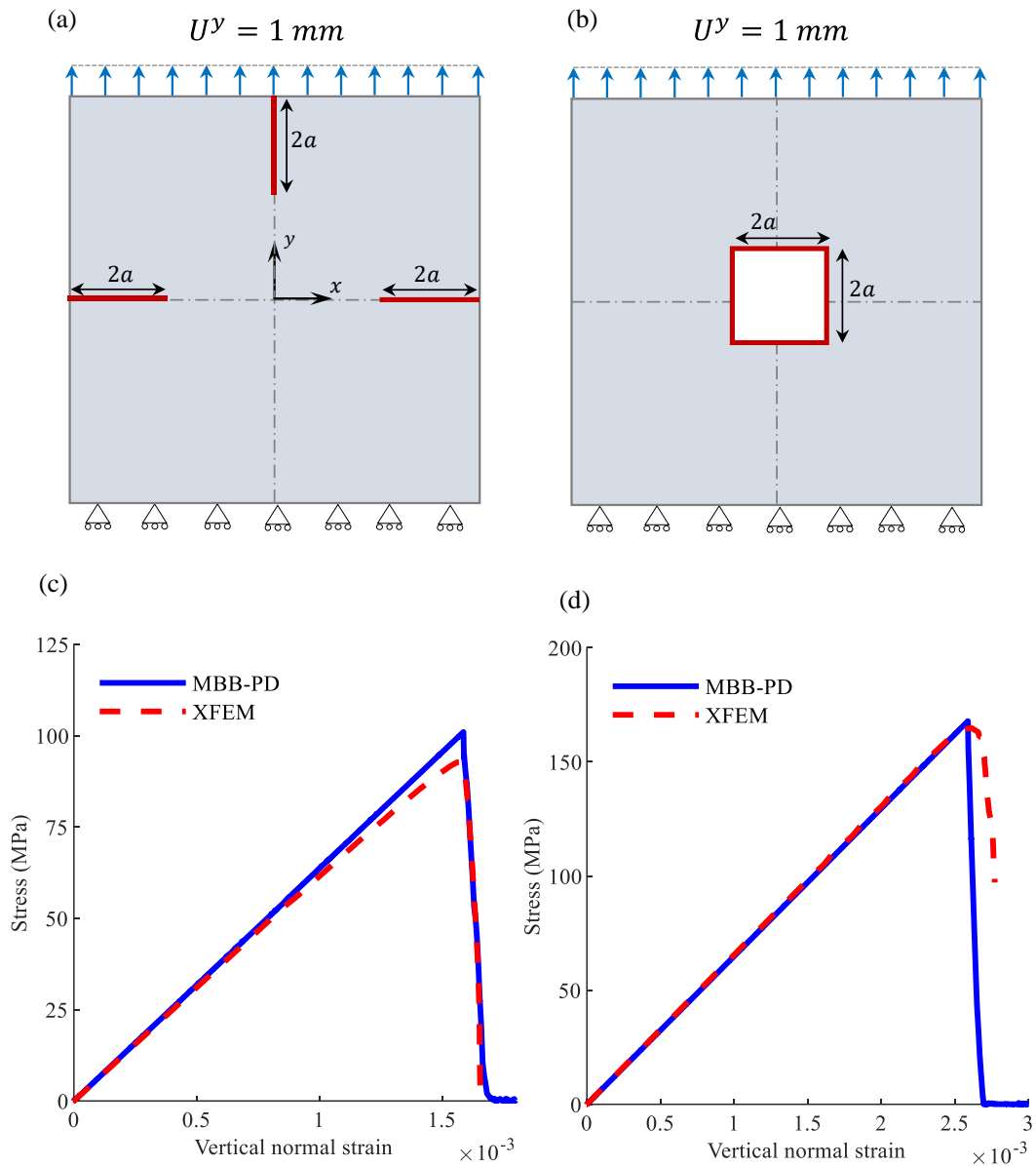

Figure 15- Schematic figures and stress-strain curves of two studied failure problems, (a) schematic figure of the problem with pre-existing three cracks, (b) schematic figure of the problem with a central square perforation, and (c), (d) comparison of the MBB-PD and XFEM simulations stress-strain curves for two problems of the (a) and (b), respectively.

The stress-strain responses of the MBB-PD model compared with the XFEM predictions for both examples are given in Figure 15(c) and (d). It can be seen that the stress-strain curves of MBB-PD and XFEM are in good agreement, with slight deviations at the end of the linear response of the multiple-crack problem. This difference can be attributed to the delay in the crack initiation predicted by the MBB-PD model compared to the XFEM analysis. This issue is more pronounced in the plate with edge cracks. In other words, the crack growth in the XFEM analysis starts earlier than the MBB-PD prediction. This is associated with the inability of the current PD model to capture progressive failure in materials. Hence, the PD models usually are appropriate for investigating the damage initiation and growth of the quasi-brittle materials, e.g., composites.

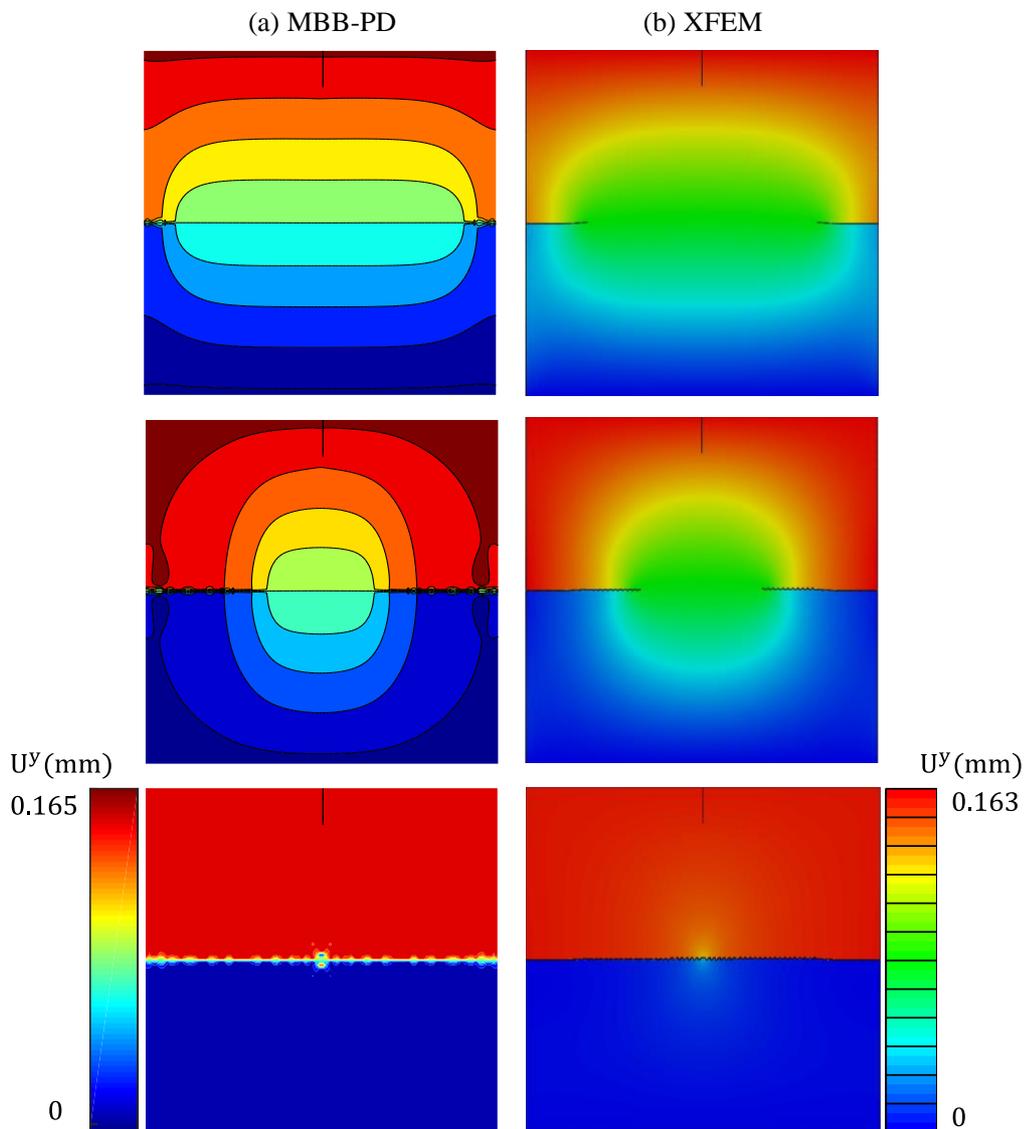

Figure 16- Vertical deformation contours of the plate with three pre-existing cracks in various time steps, (a) MBB-PD and (b) XFEM.

The vertical displacement contours of the plate with three pre-existing edge cracks simulated based on the MBB-PD and XFEM in various time steps are compared in Figure

16. It is evident that the MBB-PD model matches the XFEM simulation results in terms of crack path and deformation with a high degree of accuracy.

The damage contours of the MBB-PD simulation in various time steps are manifested and compared with XFEM in the corresponding damage states in Figure 17. An excellent agreement can be observed between the predicted damages by both methods. It can be observed that the cracks in both models nucleate from similar locations below the square corners. Furthermore, both models predict a curved-shaped damage pattern after the nucleation points. The damage grows symmetrically about the $x$ and $y$ axes until the complete separation in both models.

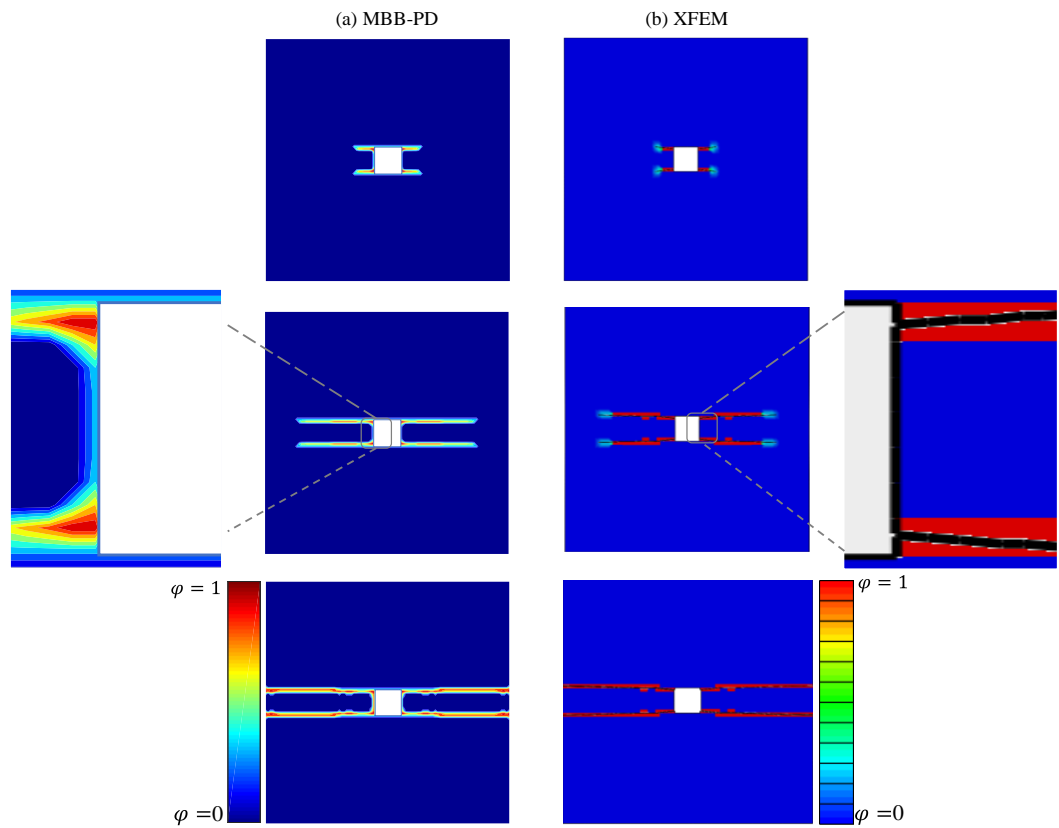

Figure 17- Damage contours of (a) MBB-PD modeling of the plate with a square hole in various time steps and (b) XFEM modeling in the corresponding damage states.

## 5. Conclusion

This study presents an MBB-PD model to remove the Poisson's ratio restriction of the classical BB-PD and increase the accuracy of the model prediction. This modification is applied by separating bonds to the horizontal/vertical normal and the shear bonds. A local imaginary element is considered for each bond to aid in estimating the bond strains after deformation at each time step. Furthermore, the true strains are considered instead of engineering strains to increase the accuracy of the numerical results. The constitutive relations are derived based on equating the strain energies of the proposed PD and CCM for a generalized combined loading condition. A critical stretch criterion and critical angle criterion are employed to predict the failure of normal and shear strain bonds, respectively. The validity of the proposed model was investigated by presenting several intact plane stress and plane strain problems. After verifying the model, the defects nucleation and growth results were presented for some pre-existing defects problems. These verifications and problems lead to the following conclusions with respect to the potential of the MBB-PD model:

(1) The MBB-PD model can overcome the limitation of the fixed Poisson's ratios in the plane stress and plane strain problems. It also conserves the simplicity and numerical cost-effectiveness of the BB-PD model.

(2) The proposed model successfully captured the elastic behavior of structures without any limitations on elastic properties under various mechanical loading (e.g., tension, tensile displacement, shear, and pressure), thermal loading, and boundary conditions. In problems with pre-existing defects, the elastic behavior of the structures until damage initiation and progression was well captured.

(3) The proposed model can predict the failure and damage growth patterns in central and edge cracks problems. Also, the MBB-PD model can precisely predict the damage initiation location and curved-shaped damage patterns in the square hole problem.

All the results of the MBB-PD were verified with XFEM results. Only slight differences were observed in the linear region before load drop. This discrepancy between the MBB-PD and XFEM results was attributed to the lack of progressive failure of bonds in the PD model. However, the proposed model can be employed to precisely predict the damage initiation and growth of the quasi-brittle materials, e.g., composite materials.

## 6. Funding

The authors received no financial support for the research, authorship and/or publication of this article.

## 7. Conflicts of interest

The authors declared no potential conflicts of interest with respect to the research, authorship, and/or publication of this article.

## 8. Acknowledgments

The authors would like to thank Mr. Amirreza Moradi for his invaluable help in FEM and XFEM simulations of problems.

## Appendix A: Modified Bond-Based Micro-modulus matrix

The micro-modulus matrix components are obtained by a combined loading condition modeling in the CCM and PD frameworks. The calibration method is based on matching the strain energy densities calculated from each framework [23,68]. The following combined loading condition is considered:

$$\varepsilon_{xx} = \varepsilon_{yy} = \gamma_{xy} = \zeta \tag{A.1}$$

Calculation of Strain Energy in the CCM framework:

The stress-strain relationship in 2D generalized elasticity is defined as:

$$\begin{Bmatrix}\sigma_{xx}\\ \sigma_{yy}\\ \sigma_{xy}\end{Bmatrix} = \begin{bmatrix}c_{11} & c_{12} & c_{16}\\ c_{12} & c_{22} & c_{26}\\ c_{16} & c_{26} & c_{66}\end{bmatrix}\begin{Bmatrix}\varepsilon_{xx}\\ \varepsilon_{yy}\\ \varepsilon_{xy}\end{Bmatrix} \tag{A.2}$$

where $\sigma_{ij}$ and $\varepsilon_{ij}$ are the in-plane stress and strain components, respectively, and $[c_{ij}]$ is the material stiffness matrix components, which is a function of the material's engineering constants. The strain Energy Density (SED) at material point $j$ in the CCM framework is defined as:

$$W_{CCM(j)} = \frac{1}{2}\frac{\int \sigma \varepsilon dV}{V} \tag{A.3}$$

By replacing Eq.(A.2) in Eq.(A.3), the generalized SED relation is obtained as:

$$W_{CCM(j)} = \frac{1}{2}(c_{11} + 2c_{12} + 2c_{16} + 2c_{26} + c_{22} + c_{66})\zeta^2 \tag{A.4}$$

Calculation of Strain Energy in the PD framework:

According to Figure A. 1, the strain-displacement relations are obtained as:

$$u_{(k)} - u_{(k)} = \varepsilon_{xx}\xi_x + \frac{\gamma_{xy}}{2}\xi_y$$

$$v_{(k)} - v_{(k)} = \frac{\gamma_{xy}}{2}\xi_x + \varepsilon_{yy}\xi_y \tag{A.5}$$

Thus, the strain components of the local elements yield:

$$S_{xx}^{(\alpha=H,V,S)} = \frac{du}{dx} = \varepsilon_{xx} + \frac{\gamma_{xy}}{2}\tan\theta_{jk}$$

$$S_{yy}^{(\alpha=H,V,S)} = \frac{dv}{dy} = \varepsilon_{yy} + \frac{\gamma_{xy}}{2}\cot\theta_{jk}$$

$$S_{xy}^{(\alpha=H,V,S)} = \frac{du}{dy} + \frac{dv}{dx} = \gamma_{xy} + \varepsilon_{yy}\tan\theta_{jk} + \varepsilon_{xx}\cot\theta_{jk} \tag{A.6}$$

Therefore, the SED under the combined loading condition using the PD framework can be evaluated as:

$$W_{PD(j)} = \frac{1}{4}\sum_{k=1}^{N_{(j)}^V+N_{(j)}^S+N_{(j)}^H}\{\vec{f}_\beta^\alpha\}\{\vec{S}_\beta^\alpha\}V_{(j)}|\vec{\xi}_{jk}| \tag{A.7}$$

where $\vec{f}_\beta^\alpha$, $\vec{S}_\beta^\alpha$, $\beta \in \{xx, yy, xy\}$ is the bond force density and the true strain of an $\alpha$-equivalent strain bond in the $\beta$-direction, respectively. Furthermore, $\alpha$ indicates the vertical, horizontal, and shear bonds. $N_{(j)}^S$, $N_{(j)}^V$, and $N_{(j)}^H$ are the number of the equivalent shear bonds, the number of the equivalent vertical bonds, and the number of the equivalent horizontal bonds of the material point of $j$.

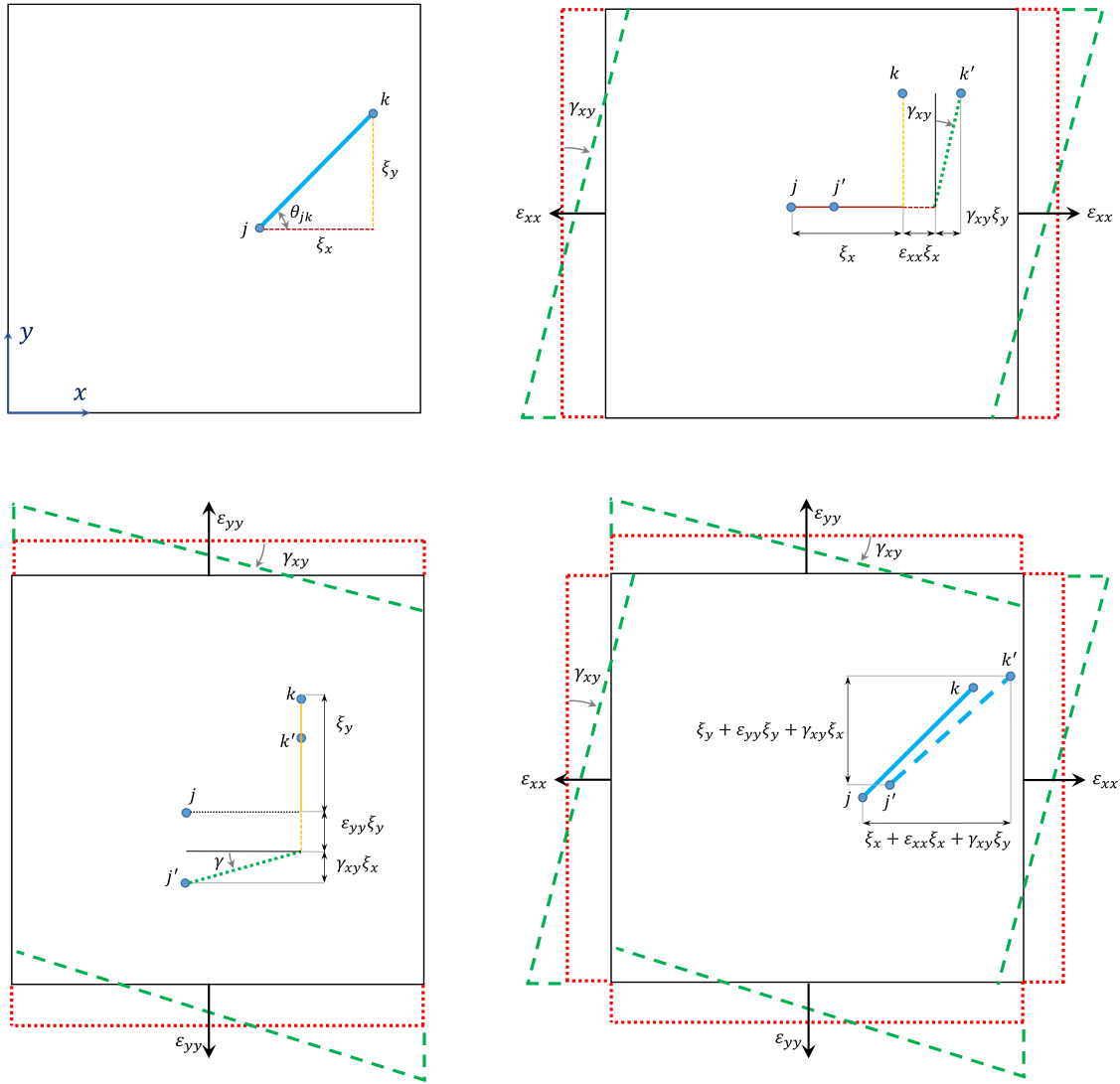

Figure A. 1- The relative position of material points.

The micro-modulus matrix is considered as a multiplication of elasticity stiffness matrix by a geometrical coefficient matrix to incorporate the material elastic constants into the micro-modulus matrix, as:

$$W_{PD(j)}$$

$$= \frac{1}{4} \sum_{k=1}^{N_{(j)}^V + N_{(j)}^S + N_{(j)}^H} \begin{bmatrix} c_{11} & c_{12} & c_{16} \\ c_{12} & c_{22} & c_{26} \\ c_{16} & c_{26} & c_{66} \end{bmatrix} \begin{bmatrix} \psi_{11} & \psi_{12} & \psi_{16} \\ \psi_{21} & \psi_{22} & \psi_{26} \\ \psi_{61} & \psi_{62} & \psi_{66} \end{bmatrix} \begin{Bmatrix} S_{xx}^{\alpha} \\ S_{yy}^{\alpha} \\ S_{xy}^{\alpha} \end{Bmatrix}^2 V_{(j)} |\vec{\xi}_{jk}| \quad (A.8)$$

Replacing Eq.(A.6) in Eq.(A.8) results:

$$W_{PD(j)}$$

$$= \frac{\zeta^2}{4} \left\{ \sum_{k=1}^{N_{(j)}^H} \psi_{11} c_1{}^t |\vec{\zeta}_{jk}| + \sum_{k=1}^{N_{(j)}^V} \psi_{22} c_2{}^t |\vec{\zeta}_{jk}| \right.$$

$$\left. + \sum_{k=1}^{N_{(j)}^S} \left( \psi_{11} c_1{}^t \left(1 + \frac{\tan \theta_{jk}}{2}\right)^2 + \psi_{22} c_2{}^t \left(1 + \frac{\cot \theta_{jk}}{2}\right)^2 + \psi_{66} c_6{}^t (\csc 2\theta_{jk} + 1)^2 \right) |\vec{\zeta}_{jk}| \right\} V_{(j)} \quad (A.9)$$

where $c_1{}^t$, $c_2{}^t$ and $c_6{}^t$ are the summation of elements of the first, second, and third row of the material stiffness matrix, respectively. By equating Eq.(A.9) and Eq.(A.4), and after some algebraic simplification, the elements of the matrix $[\psi_{jk}]$ are obtained as:

$$\psi_{11} = \psi_{12} = \psi_{16} = 2V_{(j)} \left( \sum_{k=1}^{N_{(j)}^H} |\vec{\xi}_{jk}| + \sum_{k=1}^{N_{(j)}^S} \left( \left\{1 + \frac{1}{2} \tan \theta_{jk}\right\}^2 |\vec{\xi}_{jk}| \right) \right)^{-1}$$

$$\psi_{21} = \psi_{22} = \psi_{26} = 2V_{(j)} \left( \sum_{k=1}^{N_{(j)}^v} |\vec{\xi}_{jk}| + \sum_{k=1}^{N_{(j)}^S} \left( \left\{1 + \frac{1}{2} \cot \theta_{jk}\right\}^2 |\vec{\xi}_{jk}| \right) \right)^{-1}$$

$$\psi_{61} = \psi_{62} = \psi_{66} = 2V_{(j)} \left( \sum_{k=1}^{N_{(j)}^S} \left( \{\cot \theta_{jk} + \tan \theta_{jk} + 1\}^2 |\vec{\xi}_{jk}| \right) \right)^{-1} \quad (A.10)$$

As shown in Figure A. 2, the number of horizontal or vertical bonds for the material points with the complete horizon is $2m$, and the summation of bond lengths is $m(m+1)\Delta x$ and $m(m+1)\Delta y$, respectively. Thus, Eq.(A.10) can be expressed as:

$$\psi_{11} = \psi_{12} = \psi_{16}$$

$$= 2V_{(j)} \left( m(m+1)\Delta x + \sum_{k=1}^{N_{(j)}^S} \left( \left(1 + \frac{1}{2}\tan\theta_{jk}\right)^2 |\vec{\xi}_{jk}| \right) \right)^{-1}$$

$$\psi_{21} = \psi_{22} = \psi_{26}$$

$$= 2V_{(j)} \left( m(m+1)\Delta y + \sum_{k=1}^{N_{(j)}^S} \left( \left(1 + \frac{1}{2}\cot\theta_{jk}\right)^2 |\vec{\xi}_{jk}| \right) \right)^{-1}$$

$$\psi_{61} = \psi_{62} = \psi_{66} = 2V_{(j)} \left( \sum_{k=1}^{N_{(j)}^S} \left( (\csc 2\theta_{jk} + 1)^2 |\vec{\xi}_{jk}| \right) \right)^{-1} \quad (A.11)$$

In the case of the incomplete horizon (the material points close to boundaries), $N_{(j)}^\alpha$ and $\sum_{k=1}^{N_{(j)}^\alpha} |\vec{\xi}_{jk}|$ need to be modified, where $\alpha$ represents the vertical, horizontal, and shear bonds in these relations. The modification procedure of incomplete horizons has been given in Appendix B.

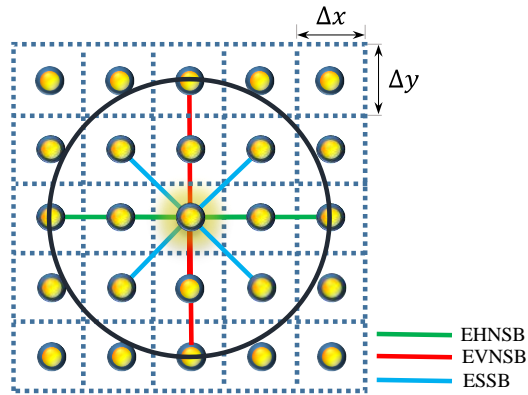

Figure A. 2- The types of bonds in a material point with $m = 2$.

## Appendix B: Volume correction factors

The geometrical coefficients for the material points with a complete horizon were derived in the previous appendix. However, these coefficients for the material points with the incomplete horizon are different (i.e., the material points near the boundaries). Therefore, some correction factors must be considered to modify the geometrical coefficients.

### B.1- EHNSB's correction factors for free surfaces

For a material point with a complete horizon in the horizontal direction, there is symmetry between the left and right sides of the point, as illustrated in Figure B. 1(b). However, this symmetry is lost if the material point is close enough to the left or right boundary. The summation of EHNSB's lengths can be written in the following form by considering the effect of the missing bonds:

$$\sum_{k=1}^{N_{(j)}^H} |\vec{\xi}_{jk}| = \begin{cases} \dfrac{m(m+1)}{2} + \left[\dfrac{REM_{(j)}(REM_{(j)}+1)}{2}\right]\Delta x & if\ REM_{(j)}^H < m \\ m(m+1)\Delta x & if\ REM_{(j)}^H \geq m \end{cases} \quad (B.1)$$

where $REM_{(j)}^H$ is the number of existing horizontal bonds on the left or right side of the material point $j$. Furthermore, the value of $N_{(j)}^H$ is rewritten as:

$$N_{(j)}^H = \begin{cases} m + REM_{(j)}^H & if\ REM_{(j)}^H < m \\ 2m & if\ REM_{(j)}^H \geq m \end{cases} \quad (B.2)$$

### B.2- EVNSBs correction factors for free surfaces

As shown in Figure B. 1(c), the correction factors for the missing vertical bonds are similar to those for the horizontal bonds, except the left and right boundaries are replaced with the bottom and top boundaries. The summation of EVNSBs' lengths with considering the effect of the missing bonds can be rewritten in the following form:

$$\sum_{k=1}^{N_{(j)}^V} |\vec{\xi}_{jk}| = \begin{cases} \dfrac{m(m+1)}{2} + \left[\dfrac{REM_{(j)}(REM_{(j)}+1)}{2}\right]\Delta y & if\ REM_{(j)}^V < m \\ m(m+1)\Delta y & if\ REM_{(j)}^V \geq m \end{cases} \quad \text{(B. 3)}$$

where $REM_{(j)}^V$ is the number of existing vertical bonds on the top or bottom side of the material point $j$. Furthermore, the value of $N_{(j)}^V$ is considered as:

$$N_{(j)}^H = \begin{cases} m + REM_{(j)}^V & if\ REM_{(j)}^V < m \\ 2m & if\ REM_{(j)}^V \geq m \end{cases} \quad \text{(B. 4)}$$

**B.3- Corrected number of ESSBs**

The shear bonds have significant roles in the material parameters of the other bonds in a horizon. Hence, identifying the correct number of the shear bonds in a horizon is important in the proposed method. The number of the shear bonds in a complete horizon, $N_{(j)}^S$, is:

$$N_{(j)}^S = NF_{(j)} - \left(N_{(j)}^V + N_{(j)}^H\right) \quad \text{(B. 5)}$$

where $NF_{(j)}$, $N_{(j)}^V$, and $N_{(j)}^H$ are the total number of equivalent bonds, the number of the equivalent vertical bonds, and the number of the equivalent horizontal bonds of the material point of $j$, respectively. The missing bonds, which are close to the boundaries and must be ignored in relations, are shown in Figure B. 1(d).

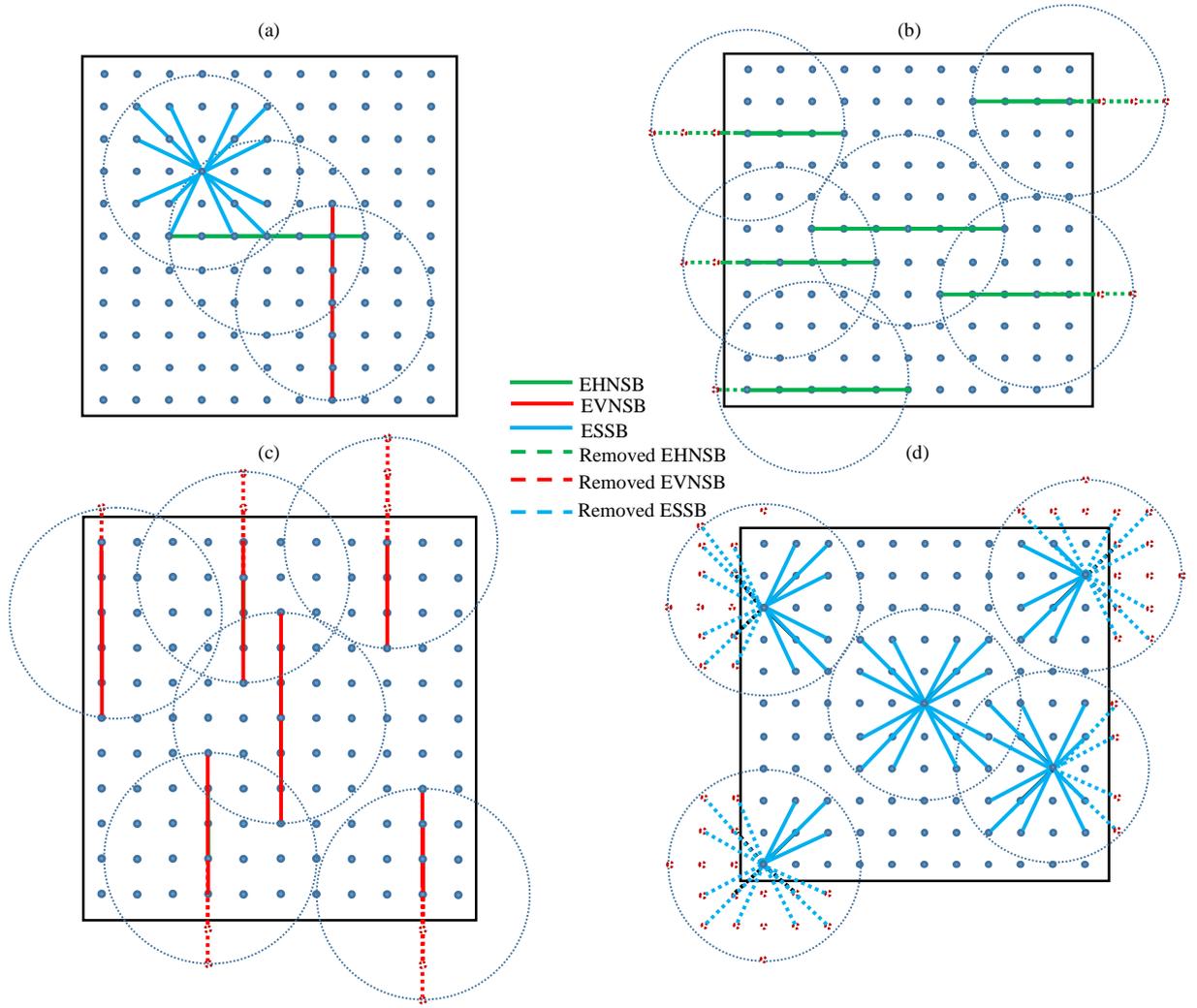

Figure B. 1- The lost bonds and corrections in various locations and bond situations, (a) the families of the horizontal, vertical, and shear bonds far from free surfaces, (b) the families and the lost horizontal bonds near the free surfaces, (c) the families and the lost vertical bonds near the free surfaces, (d) the families and the lost shear bonds near the free surfaces.

## Appendix C: Critical stretch and critical angle for bond's failure

Two types of failure criteria are generally applied in PD models: the distributed SERR criterion [69] and the critical stretch criterion [70]. The energy is divided by the number of bonds that pass across a unit crack area in the SERR criterion. This model considers a uniform energy distribution over the bonds, regardless of the bonds' geometrical characteristics. However, a critical stretch value is proposed for all bonds in the critical stretch criterion, regardless of bond type. The shear bonds have the bond angle deformation dominantly, and the stretch of the bonds is not significant. So, for more accurate modeling of the failures, it is essential to consider a critical value for the angle of the shear bonds.

Mode I (opening Mode), Mode II (sliding Mode), and Mixed Mode of fracture are considered in the current study. In each failure mode, both the normal and shear bonds are involved. Therefore, different critical values must be computed for the normal stretches and the shear angles.

**C.1- Mode I**

As discussed earlier, the critical SERR is the total energy required for breaking all the interactions of a material point passing through the crack surface (as shown in Figure C. 1(a)). Thus, the SERR of Mode I for a crack surface with a length of $\Delta x$ can be written as:

$$G_{I(j)} = \sum_{k=1}^{N_{cr}^V + N_{cr}^S} \frac{\mu(\{\vec{f}_\beta^\alpha\}\{\vec{S}_\beta^\alpha\} V_{(j)} V_{(k)} \vec{\xi}_{jk}}{2(\Delta x) h} \tag{C.1}$$

where $\beta \in \{xx, yy, xy\}$ and $\alpha$ indicate the vertical, horizontal, and shear bonds. In Mode I, both the normal and the shear interactions undergo deformation. Therefore, the corresponding critical SERR of these interactions is obtained as:

$$G_{I(j)}^c = \sum_{k=1}^{N_{cr}^S} \frac{\mu[K_{jk}]\begin{Bmatrix} -v_{xy}s_c \\ s_c \\ s_c \sin\theta_{jk} \end{Bmatrix}^2 V_{(j)}V_{(k)}\vec{\xi}_{jk}}{2(\Delta x)h}$$

$$+ \sum_{k=1}^{N_{cr}^V} \frac{\mu[K_{jk}]\begin{Bmatrix} -v_{xy}s_c \\ s_c \\ 0 \end{Bmatrix}^2 V_{(j)}V_{(k)}\vec{\xi}_{jk}}{2(\Delta x)h} \quad (C.2)$$

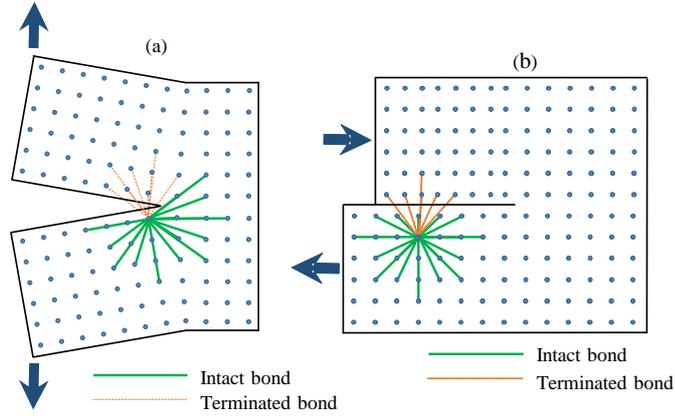

Figure C. 1- Material points' interactions in various failure modes, (a) failure Mode I and (b) failure Mode II.

According to Figure C. 2(a), and using trigonometry, the relationship between the critical stretch ($s_c^I$) and the critical angle ($\phi^I_c$) is obtained as:

$$\phi^I_{c(jk)} = s_{c(j)}^I \sin\theta_{jk} \quad (C.3)$$

By substituting the material parameters in Eq.(C.2), the critical stretch yields:

$$s_{c(j)}^I = \left(\left(\left(\sum_{k=1}^{N_{cr}^V} \mu\left\{K_2^t \vec{j}\right\} \vec{\xi}_{jk}\right.\right.\right.$$

$$\left.+ \sum_{k=1}^{N_{cr}^S} \mu\left\{\left(v_{xy}^2 K_1^t + \left(\frac{1}{2}\tan\theta_{jk}\sin^2\theta_{jk}\right)K_6^t\right)\vec{i}\right.\right.$$

$$\left.\left.\left.+ \left(K_2^t + \left(\frac{1}{2}\cot\theta_{jk}\sin^2\theta_{jk}\right)K_6^t\right)\vec{j}\right\}\vec{\xi}_{jk}\right)^{-1} 2G_I^C \Delta x h V^{-2}\right)^{\frac{1}{2}}$$ (C.4)

**C.2-Mode II**

The Bonds' configurations in fracture Mode II are illustrated in Figure C. 1(b). As can be seen, both the normal and shear bonds fail during the Mode II fracture. Similar to Eq.(C.2), the critical SERR of Mode II for a crack surface with a length of $\Delta x$ can be written as:

$$G_{II(j)}^c = \sum_{k=1}^{N_{cr}^S + N_{cr}^V} \frac{\mu[K_{jk}]\begin{Bmatrix} s_c \tan\theta_{jk} \\ 0 \\ s_c \end{Bmatrix}^2 V_{(j)}V_{(k)}\vec{\xi}_{jk}}{2(\Delta x)h}$$ (C.5)

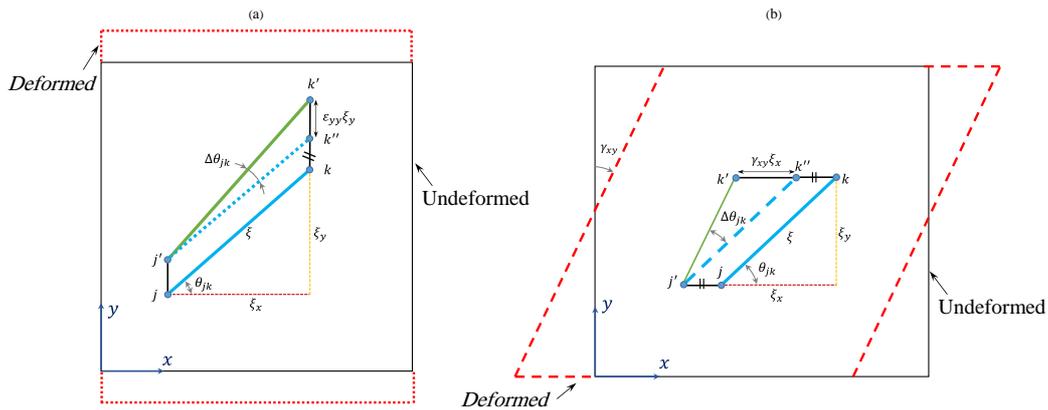

Figure C. 2- Shear bonds deformations and angle variations in the various failure modes, (a) Mode I , (b) Mode II.

Also, according to Figure C. 2(b), the relationship between the critical stretch ($s_c^{II}$) and the critical angle ($\phi^{II}{}_c$) is obtained as:

$$\phi^{II}{}_{c(jk)} = s^{II}_{c(j)} \tag{C.6}$$

After that, the critical stretch under Mode II configuration is obtained as:

$$s^{II}_{c(j)} = \left(\left(\left(\sum_{k=1}^{N^V_{cr}} \mu \left\{\frac{1}{2}\cot\theta_{jk} K_2{}^t\vec{j}\right\}\vec{\xi}_{jk}\right.\right.\right.$$

$$+ \sum_{k=1}^{N^S_{cr}} \mu \left\{\left(\frac{1}{2}\tan\theta_{jk} K_6{}^t + K_1{}^t \tan\theta_{jk}{}^2\right)\vec{\imath}\right.$$

$$\left.\left.\left.+ \frac{1}{2}\cot\theta_{jk} K_6{}^t\vec{j}\right\}\vec{\xi}_{jk}\right)^{-1} 2G^C_{II}\Delta x h V^{-2}\right)^{\frac{1}{2}} \tag{C.7}$$

**C.3. Mixed-Mode**

As mentioned in section 3.3, the B-K criterion is used in the case of the mixed-Mode fracture. Therefore, under a mixed-mode loading condition, the mixed-Mode critical SERR for a crack surface with a length of $\Delta x$ can be expressed as:

$$G^C_{M(j)} = \sum_{k=1}^{N^V_{cr}+N^S_{cr}} \frac{[K_{jk}]\begin{Bmatrix} s_c \tan\theta_{jk} - \nu_{xy}s_c \\ s_c \\ s_c + s_c \sin\theta_{jk} \end{Bmatrix}^2 V_{(j)}V_{(k)}\vec{\xi}_{jk}}{2(\Delta x)h} \tag{C.8}$$

Consequently, the critical stretch for a mixed-mode loading is obtained as:

$$s_{c(j)}^{M} = \left(\left(\sum_{k=1}^{N_{cr}^{V}} \mu\left\{\left(K_1{}^t + \frac{1}{2}\cot\theta_{jk} K_6{}^t\right)\vec{j}\right\}\vec{\xi}_{jk}\right.\right.$$

$$+ \sum_{k=1}^{N_{cr}^{S}} \mu\left\{\left((\tan\theta_{jk} - v_{xy})^2 K_1{}^t + \frac{1}{2}\tan\theta_{jk}(1+\sin\theta_{jk})^2 K_6{}^t\right)\vec{\imath}\right.$$

$$\left.\left.+ \left(K_2{}^t\vec{j} + \frac{1}{2}\cot\theta_{jk}(1+\sin\theta_{jk})^2 K_6{}^t\right)\vec{j}\right\}\vec{\xi}_{jk}\right)^{-1} 2G_M^C \Delta x h V^{-2}\right)^{\frac{1}{2}} \quad (C.9)$$

The relationship between angle and the mixed-mode loading condition is indicated in Figure A. 1. According to the relationship of the deformations in Eq.(A.5), the critical angle is given as:

$$\phi_{c(jk)}{}^M = s_{c(j)}^{II}(\pm 1 \pm \sin\theta_{jk}) \quad (C.10)$$